\documentclass[draft]{agujournal2019}
\usepackage{url} 
\usepackage{soul}
\usepackage[load-configurations=abbreviations]{siunitx}
\usepackage{amsfonts,amsmath,amssymb,amsthm}
\usepackage{bm}
\usepackage{booktabs} 

\draftfalse

\journalname{AGU Advances}

\usepackage[final]{pdfpages} 

\begin{document}

%
%


\title{Europa's ocean translates interior tidal heating patterns to the ice-ocean boundary}

%
%




\authors{D. G. Lemasquerier\affil{1,2}, C. J. Bierson\affil{3}, K. M. Soderlund\affil{1}}


\affiliation{1}{University of Texas at Austin, Jackson School of Geosciences, Institute for Geophysics, Austin, United States}
\affiliation{2}{University of St Andrews, School of Mathematics and Statistics, St Andrews, United Kingdom}
\affiliation{3}{Arizona State University, School of Earth and Space Exploration, Tempe, United States}




\correspondingauthor{Daphné Lemasquerier}{d.lemasquerier@st-andrews.ac.uk}



\begin{keypoints}
\item We use an idealized model of thermally-driven flows in Europa's ocean, neglecting salinity and feedback effects of the ice
\item Heterogeneous tidal heating in the mantle modifies the mean circulation in Europa's ocean and could drive large-scale thermal winds
\item The tidal heating anomaly in latitude is efficiently translated upwards, leading to a higher heat flux into the ice shell at the poles
\end{keypoints}

%
%

%
%


\begin{abstract}
	The circulation in Europa's ocean determines the degree of thermal, mechanical and chemical coupling between the ice shell and the silicate mantle. Using global direct numerical simulations, we investigate the effect of heterogeneous tidal heating in the silicate mantle on rotating thermal convection in the ocean and its consequences on ice shell thickness. Under the assumption of no salinity or ocean-ice shell feedbacks, we show that convection largely transposes the latitudinal variations of tidal heating from the seafloor to the ice, leading to a higher oceanic heat flux in polar regions. Longitudinal variations are efficiently transferred when boundary-driven thermal winds develop, but are reduced in the presence of strong zonal flows {and may vanish in planetary regimes}. If spatially homogeneous radiogenic heating is dominant in the silicate mantle, the ocean's contribution to ice shell thickness variations is negligible compared to tidal heating within the ice. If tidal heating is instead dominant in the mantle, the situation is reversed and the ocean controls the pole-to-equator thickness contrast, as well as possible longitudinal variations.  
\end{abstract}

\section*{Plain Language Summary}
Europa, an icy moon of Jupiter, is believed to have a deep salty ocean beneath its ice crust. One of the drivers of ocean circulation is heating from the rocky mantle located under the ocean. This heating is due to 1) the decay of radioactive elements in the mantle (``radiogenic heating''), and 2) the periodic deformation of the mantle as Europa revolves around Jupiter, due to the gravitational force exerted by the gas giant (``tidal heating''). Tidal heating is strongly heterogeneous: higher at the poles, and lower at the points facing and opposite to Jupiter. We investigate the effect of large-scale heating variations using simulations of the ocean dynamics, although at less extreme parameters than the real Europa ocean, and neglecting the effects of salinity and phase change. We show that if tidal heating is dominant, the ocean circulation does not erase the variations of bottom heating and transposes them particularly well in latitude up to the ice-ocean boundary. This has consequences on the ice shell equilibrium: if mantle heating is heterogeneous, thickness variations could be controlled by the oceanic heat flux, resulting in thinner ice at the poles. These results now await comparison with measurements from \textit{Europa Clipper}.

%
%

\section{Introduction}

Among icy ocean worlds, Europa has one of the best potentials to harbour habitable environments. Unlike Ganymede or Titan \cite{vance_geophysical_2018}, Europa's ocean \cite{kivelson_galileo_2000} is in direct contact with the mantle rocks \cite{anderson_europas_1998}. This Earth-like configuration could promote contact between reductants from water-rock interactions with oxidants produced at the ice's surface by irradiation, and provide energy to sustain life \cite{hendrix_nasa_2019,howell_nasas_2020}.
The ocean also controls the interior structure and evolution of the moon by coupling the silicate mantle with the ice crust \cite{soderlund_ice-ocean_2020}. However, its composition and dynamics remain vastly unknown. The upcoming \textit{Europa Clipper} mission \cite{phillips_europa_2014,howell_nasas_2020,roberts2023exploring} will provide new constraints on the moon's interior, but the ocean dynamics can only be indirectly characterized. Forward dynamical models are necessary to determine the spatio-temporal properties of heat and material exchanges between the ice and the silicate interior, and guide future interpretation.

The flows in Europa's ocean can be buoyancy-driven by thermal or compositional gradients, mechanically-driven by tides or libration, or electromagnetically-driven by the Jovian magnetic field \cite<see>[for a review, and references therein]{soderlund_physical_2024}. Here, we focus on thermally-driven flows to characterize the heat exchange between the ice shell and the deep interior. {We, therefore, caution that the circulation patterns obtained in our study neglect salinity effects as well as those due to phase changes at the ice-ocean boundary.} \citeA{soderlund_ocean-driven_2014} suggested that the heat flux from Europa's ocean is higher at the equator, thereby melting and thinning the ice at low latitudes, a configuration called \textit{equatorial cooling}. However, this work and following studies \cite{soderlund_ocean_2019,amit_cooling_2020,kvorka_numerical_2022} considered Europa's ocean to be uniformly heated by the silicate mantle, which is possibly unrealistic if tidal heating exceeds radiogenic heating in the mantle.

\begin{figure}[ht]
	\centering
	\includegraphics[width=1\linewidth]{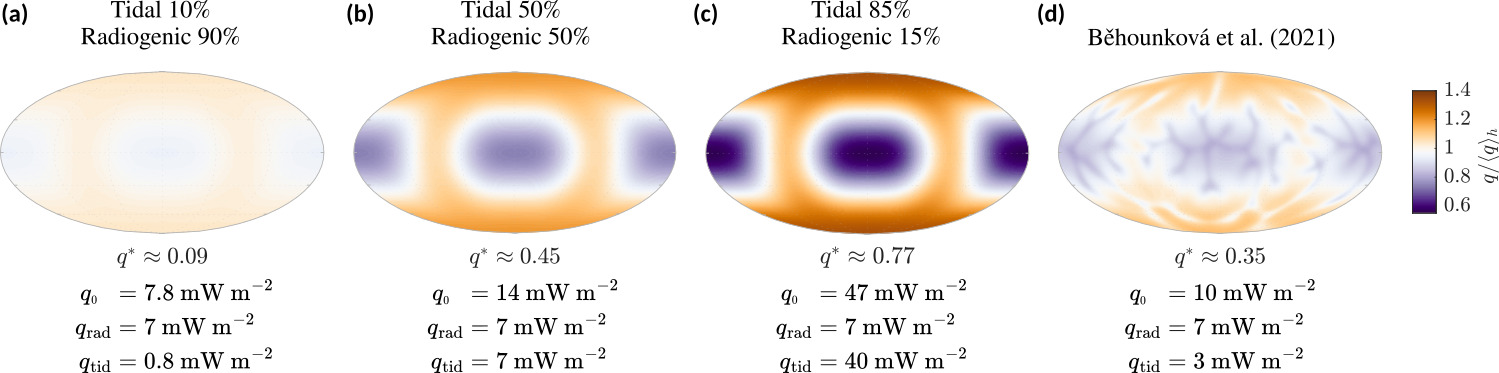}
	\caption{Patterns of heat flux per unit area at the seafloor in various scenarios. $q^* = \Delta q/q_0$, the ratio between the maximum heat flux difference and the laterally averaged heat flux, measures the heterogeneity of the heat flux. \textbf{(a)} If radiogenic heating is dominant in the silicate mantle, heating of the ocean is homogeneous ($q^*\approx 0$). \textbf{(b)} As the relative amplitude of tidal heating is increased, heat flux from the mantle becomes heterogeneous. For dominant tidal heating \textbf{(c)} relative heat flux variations become of order one ($q^*\rightarrow 0.91$). The tidal heating pattern is computed using the tidal model of \citeA{roberts_tidal_2008} described in Supplementary Text S1.  \textbf{(d)} Heat flux pattern obtained at the end of Europa's thermal evolution model by  \citeA{behounkova_tidally_2021} : the tidal heating represents about 30\% of the total heating in the silicate mantle. Despite the occurrence of melting and mantle convection, the large-scale tidal heating pattern is persistent. Figure modified from Figure 2(d) in \citeA{behounkova_tidally_2021}, using the corresponding dataset \cite{marie_behounkova_2020_3950431} licensed under CC BY 4.0 (https://creativecommons.org/licenses/by/4.0/).}
	\label{fig:tidalheating}
\end{figure}

Unlike radiogenic heating, tidal heating is spatially heterogeneous \cite<e.g., >{tobie_tidal_2005,sotin_tides_2009,beuthe_spatial_2013}. For Europa's silicate mantle, it is predicted to be significantly greater at the poles than at the equator, with longitudinal variations of order 2 at low latitudes as well  (Figure \ref{fig:tidalheating} and Supplementary Figure S2). To measure how strongly heterogeneous this pattern is, we introduce the parameter $q^* = \Delta q/q_0$, the ratio between the maximum heat flux difference (between the poles and the sub or anti-Jovian points) and the laterally averaged heat flux, $q_0$. For pure tidal heating, $q^*\sim 0.91$, i.e. heat flux variations are of the same order as the mean heat flux \cite{tobie_tidal_2005,beuthe_spatial_2013}. However, tidal heating is superposed on radiogenic heating, which is nominally homogeneous ($q^*=0$). The relative amplitude of tidal heating compared to radiogenic heating in Europa's silicate mantle is poorly constrained \cite{hussmann_constraints_2016} and has likely varied during Europa's evolution \cite{behounkova_tidally_2021}. Therefore, $q^*$ can a priori lie anywhere between 0 and 1 (Figure \ref{fig:tidalheating}) and we consider a sweep in $q^*$ to evaluate the ocean's response in various scenarios. 

How rotating thermal convection responds to a heterogeneous heating has been investigated for Earth's outer core  \cite<e.g.,>{sahoo_response_2020,mound_regional_2019,davies_mantle-induced_2019,dietrich_core_2016,olson_core_2015,gubbins_melting_2011}. It was found that the heat flux pattern at the core-mantle boundary could not reach the inner core \cite{davies_mantle-induced_2019}. However, Earth's outer core is a much thicker shell compared to Europa's ocean, and we explore less rotationally-controlled regimes compared to the geostrophic regime investigated for the Earth \cite{mound_heat_2017,davies_mantle-induced_2019}. In the context of icy moon oceans, motivated by convection patterns in high-pressure ices of Titan or Ganymede \cite{choblet_heat_2017}, it was recently showed that small scale heating patterns do not influence convection at a global scale, and that the ocean cooling configuration (polar or equatorial) is unaffected \cite{terra-nova_influence_2022}. Narrow high-heating bands at the seafloor of Enceladus would also be mixed by the ocean and only have a diffuse imprint beneath the ice \cite{kang_ocean_2022}. Large-scale heating patterns could lead to different conclusions, because large-scale temperature gradients can drive a mean thermal-wind circulation \cite{dietrich_core_2016}. In the present work, we investigate 1) how the large-scale tidal heating of Europa's mantle impacts its ocean circulation, 2) whether large-scale thermal anomalies from the seafloor can be transposed up to the ice-ocean boundary, and 3) if resulting oceanic heat flux variations could affect the ice crust equilibrium.

\section{Materials and Methods}

To address these questions, we model the ocean by a fluid contained in a rotating spherical shell and driven by 1) vertical convection due to the heating of the ocean from below, and 2) horizontal convection \cite{hughes_horizontal_2008,gayen_rotating_2022} due to lateral variations of the basal heat flux. A competition between the vertically-forced and horizontally-forced convection will take place \cite{couston_competition_2022} and determine the properties of the large-scale circulation. In the presence of rotation, thermal winds develop due to horizontal temperature gradients \cite[and references therein]{dietrich_core_2016}. To investigate this, we perform direct numerical simulations of turbulent thermal convection with an imposed heat flux at the bottom boundary. The basal heat flux is either homogeneous (representing radiogenic heating) or increasingly heterogeneous following the tidal heating pattern in the mantle. These methods are described in more detail below.

From the ocean model, we extract relative heat flux variations obtained at the ice-ocean boundary. In section \ref{sec:icethickness}, we use them as inputs in an ice thickness model {(see} {\ref{app:iceinputs}} {and Supplementary Text S2)} to evaluate the possible impact of a heterogeneous oceanic heat flux on the ice shell thermal equilibrium. {Note that this is not a coupled model, i.e. the ocean dynamics are solved separately from the ice shell equilibrium. Therefore, important feedback processes are missing (e.g., topography of the ice-ocean boundary, freezing point depression with pressure). The ice thickness maps obtained here are not meant to be definitive quantitative predictions for Europa, but are rather used as a tool to compare the relative effects of oceanic heat flux variations versus tidal heating variations within the ice.}

\subsection{Set of Equations}

We numerically model the ocean circulation by solving for turbulent thermal convection in rotating spherical shells using the open-source code MagIC \cite{wicht_inner-core_2002,christensen_numerical_2001,schaeffer_efficient_2013} (URL: https://magic-sph.github.io/). The ocean is modelled as a fluid of kinematic viscosity $\nu$, density $\rho$, thermal diffusivity $\kappa$, heat capacity $c_p$, and thermal expansivity $\alpha$,  contained in a spherical shell of outer and inner radii $r_o$ and $r_i$, thickness $D=r_o-r_i$, rotating at a {uniform} rate $\Omega$ about the vertical axis $z$. The aspect ratio of the shell, defined as $\eta = 1 - D/r_o$ is fixed at 0.9, close to the aspect ratio of Europa's ocean (Table \ref{tbl:icymoonsparams}).  
In the following, all of the variables are made dimensionless using the shell thickness $D$ for the length scale and the viscous time $D^2/\nu$ for the timescale. The temperature scale is $D  q_0  / k$, where $q_0=\langle q_{\rm bot} \rangle_h = k\langle \partial_rT \vert_{r=r_i} \rangle_h$ is the horizontal average of the imposed basal heat flux and $k=\rho c_p \kappa$ is the thermal conductivity of the fluid. In the Boussinesq approximation, and in the rotating frame of reference, the set of dimensionless equations governing the system are:
\begin{linenomath*}
	\begin{align}
		\frac{\partial \bm{u}}{\partial t} + \bm{u \cdot \nabla} \bm{u}  &= - \bm{\nabla} p - \frac{2}{E} \bm{e}_z \times \bm{u} + \frac{Ra_F}{Pr} \frac{r}{r_o} T \bm{e}_r +  \nabla^2 \bm{u}, \label{eq:NS} \\
		\frac{\partial T}{\partial t} +  \bm{u \cdot \nabla} T &= \frac{1}{Pr} \nabla^2 T, \label{eq:energy}\\
		\bm{\nabla \cdot u} &= 0, \label{eq:mass}
	\end{align}
\end{linenomath*}
where equation \eqref{eq:NS} accounts for conservation of momentum (Navier-Stokes equation), equation \eqref{eq:energy} accounts for energy conservation and equation \eqref{eq:mass} accounts for mass conservation. Here, $\bm{e}_r$ and $\bm{e}_z$ are the radial and vertical unit vectors, respectively, {where the vertical is aligned with the rotation axis}. $T$,  $p$ and $\bm{u}$ are the dimensionless fluid temperature, pressure and velocity, and the gravity field is taken to be linear $\bm{g}=g_o \bm{r}/r_o$. At a fixed aspect ratio, the solution of the system of equations \eqref{eq:NS}-\eqref{eq:mass} depends on the relative value of three dimensionless parameters: the flux Rayleigh number $Ra_F$, the Ekman number $E$, and the Prandtl number $Pr$:
\begin{linenomath*}
	\begin{equation}
		\begin{aligned}
			Ra_F &=& \dfrac{\rm buoyancy}{\rm viscous + thermal~diffusion} &=& \frac{\alpha g_o q_0 D^4 }{\rho c_p \nu \kappa^2}, \\
			E  &=& \dfrac{\rm viscous ~diffusion}{\rm rotation} &=& \frac{\nu}{\Omega D^2}, \\
			Pr &=& \dfrac{\rm viscous ~diffusion}{\rm thermal ~diffusion} &=& \frac{\nu}{\kappa}.
		\end{aligned}
	\end{equation}
\end{linenomath*}
Since we considerer a heterogeneous basal heat flux, a fourth dimensionless parameter is introduced to represent the relative basal heat flux anomaly,
\begin{linenomath*}
	\begin{equation}
		q^* = \frac{\Delta q_{\rm bot}}{q_0}.
	\end{equation}
\end{linenomath*}
Here, $\Delta q_{\rm bot}$ is the heat flux difference between the maximum and the minimum basal heat flux. Given the tidal pattern (Figure \ref{fig:tidalheating}), the maximum heat flux point is always at the pole, and the minimum heat flux is located at the sub-jovian and anti-jovian points.

\subsection{Boundary Conditions}

The top and bottom boundaries are impenetrable and either both no-slip or both stress-free. We enforce angular momentum conservation in the fluid bulk. No-slip mechanical boundary conditions could seem the most natural ones to employ, but given the artificially high viscous forces in our simulations {(and in all direct numerical simulations (DNS) of convection in a 3D shell), necessary to make the problem computationally-tractable,} no-slip conditions lead to the development of thick Ekman boundary layers which are not planetary relevant \cite<e.g.,>{kuang_earth-like_1997,glatzmaier2002geodynamo}. We therefore also consider stress-free boundary conditions. 

We use a Neumann boundary condition at the bottom of the ocean -- fixed flux $q_{\rm bot}(\theta,\phi)$ -- to represent the heat flux from the silicate mantle. We use a Dirichlet boundary condition at the top of the ocean --fixed temperature $T_{\rm top}$-- to represent the fixed melting temperature. Note that we also investigated cases where the temperature is fixed at both boundaries in order to compare with and extend previous studies mostly done with Dirichlet boundary conditions \cite{soderlund_ocean-driven_2014,soderlund_ocean_2019,amit_cooling_2020,kvorka_numerical_2022}. These Dirichlet simulations are described in the Supplementary Information.

To impose a varying heat flux pattern on the bottom boundary, the map represented in Figure \ref{fig:tidalheating}(c) is decomposed into spherical harmonics (Supplementary Figure S2). The bottom heat flux can thus be expressed as
\begin{linenomath*}
	\begin{equation*}
		q_\textup{bot}(\theta,\phi)= \sum_{\ell = 0}^{\ell_{max}} \sum_{m=-\ell}^{\ell} q_\textup{bot}^{\ell m} Y_\ell^m(\theta,\phi),
	\end{equation*}
\end{linenomath*}
where $\theta$ is the colatitude and $\phi$ the longitude. Here $\ell$ and $m$ denote the spherical harmonic degree and order respectively, and the spherical harmonics $Y_\ell^m$ are completely normalized: 
\begin{linenomath*}
	\begin{equation*}
		Y_\ell^m(\theta,\phi) = \left[ \frac{2\ell +1}{4\pi} \frac{(\ell -\vert m\vert) !}{(\ell +\vert m\vert) !} \right]^{1/2} P_\ell^m(\cos (\theta)) e^{i m\phi},
	\end{equation*}
\end{linenomath*}
with $P_\ell^m$ the associated Legendre functions. We keep only the most significant modes, and therefore truncate the decomposition at degree $\ell=4$ (Supplementary Figure S2). The corresponding spherical harmonic coefficients, $q_\textup{bot}^{\ell m}$ are provided in Table \ref{tbl:tidalpattern}. For the ocean numerical simulations, a sweep in $q^*$ is performed by keeping the average heat flux fixed and progressively increasing the amplitude of all the other coefficients $(l,m)\neq(0,0)$. The \textit{relative} amplitude of the modes $(l,m)\neq(0,0)$ between each other is always the same, so that the pattern is geometrically fixed. Coefficients for Dirichlet simulations are also provided in Supplementary Table S2.

\begin{table}[ht]
	\centering
	\resizebox{\textwidth}{!}{%
	\begin{tabular}{lcccccc}
		\multicolumn{3}{l}{\textbf{Tidal model}}  \\
		$q^*$ & $(\ell=0,m=0)$ & $(\ell=2,m=0)$ & $(\ell=2,m=2)$ & $(\ell=4,m=0)$ & $(\ell=4,m=2)$ & $(\ell=4,m=4)$ \\
		\toprule[0.5pt]
		0.91 &  1 & \num{2.10e-1} & \num{-9.92e-2} & \num{-2.44e-2} & \num{2.28e-2} & \num{2.35e-2} \\
		\toprule[0.5pt]					
		\multicolumn{4}{l}{\textbf{Ocean Simulations - imposed basal heat flux}}  \\
		$q^*$ & $(\ell=0,m=0)$ & $(\ell=2,m=0)$ & $(\ell=2,m=2)$ & $(\ell=4,m=0)$ & $(\ell=4,m=2)$ & $(\ell=4,m=4)$ \\
		\toprule[0.5pt]
		0.00 & -1.0 & 0 & 0 & 0 & 0 & 0\\
		0.46 & -1.0 &\num{-0.1} &\num{4.6e-2} &\num{1.1e-2}  &\num{-9.8e-3}  &\num{-1.02e-3}\\
		0.87 & -1.0 &\num{-0.2} &\num{9.2e-2} &\num{2.16e-2} &\num{-2.0e-2}  &\num{-2.0e-2}\\
		1.30 & -1.0 &\num{-0.3} &\num{1.4e-1} &\num{3.2e-2}  &\num{-2.9e-2}  &\num{-3.1e-2}\\
		1.72 & -1.0 &\num{-0.4} &\num{1.8e-1} & \num{4.3e-2} &\num{-3.9e-2}  & \num{-4.1e-2}\\
		2.16 & -1.0 &\num{-0.5} &\num{2.3e-1} & \num{5.4e-2} & \num{-4.9e-2} & \num{-5.11e-2} \\	
		\bottomrule
	\end{tabular}
	}
	\caption{\label{tbl:tidalpattern} Spherical harmonic coefficients used for the dimensionless basal heat flux. The coefficients for the tidal model correspond to the map of Figure \ref{fig:tidalheating}(c), with the coefficients normalized by the $(\ell=0,m=0)$ coefficient. For the ocean numerical simulations, the average heat flux is kept fixed ($q_{bot}^{l=0,m=0}$ is constant), but the amplitude of all the other coefficients $(l,m)\neq(0,0)$ is progressively increased to increase $q^*$. The \textit{relative} amplitude of the modes $(l,m)\neq(0,0)$ between each other is always the same. Note that the relative amplitudes between the coefficients (4,0), (4,2) and (4,4) compared to the mode (2,2) very slightly differ between the simulations and the tidal model because of an initial error in our tidal dissipation code. As these modes have the smallest amplitude, this does not significantly affect the tidal pattern nor the fluid's response.}
\end{table} 

\subsection{Dimensionless parameters}

In this work, the goal is to survey how different possible contributions of tidal heating in the mantle would impact ocean circulation and heat transport, not to quantify tidal heating in Europa's mantle. Instead, we consider a wide range of possible heat flux amplitudes at the seafloor, going from 6 \si{\milli\watt\per\meter\squared} to 46 \si{\milli\watt\per\meter\squared}. The lower bound corresponds to the case of pure radiogenic heating \cite{tobie_tidally_2003} and negligible tidal heating, whereas the upper bound accounts for the possibility of a dominant, Io-like tidal dissipation \cite{howell_likely_2021} (see also Supplementary Text S1).
With the other physical parameters listed in Table \ref{tbl:icymoonsparams}, we obtain for Europa's ocean $Pr\approx 11$, $E= \num{7e-12}$ ($\in[\num{5.0e-12},\num{9.1e-12}]$) and $Ra_F=\num{5.4e27}$ ($\in[\num{6.3e26},\num{2.8e28}]$).

\begin{figure}[p]
	\centering
	\includegraphics[width=\linewidth]{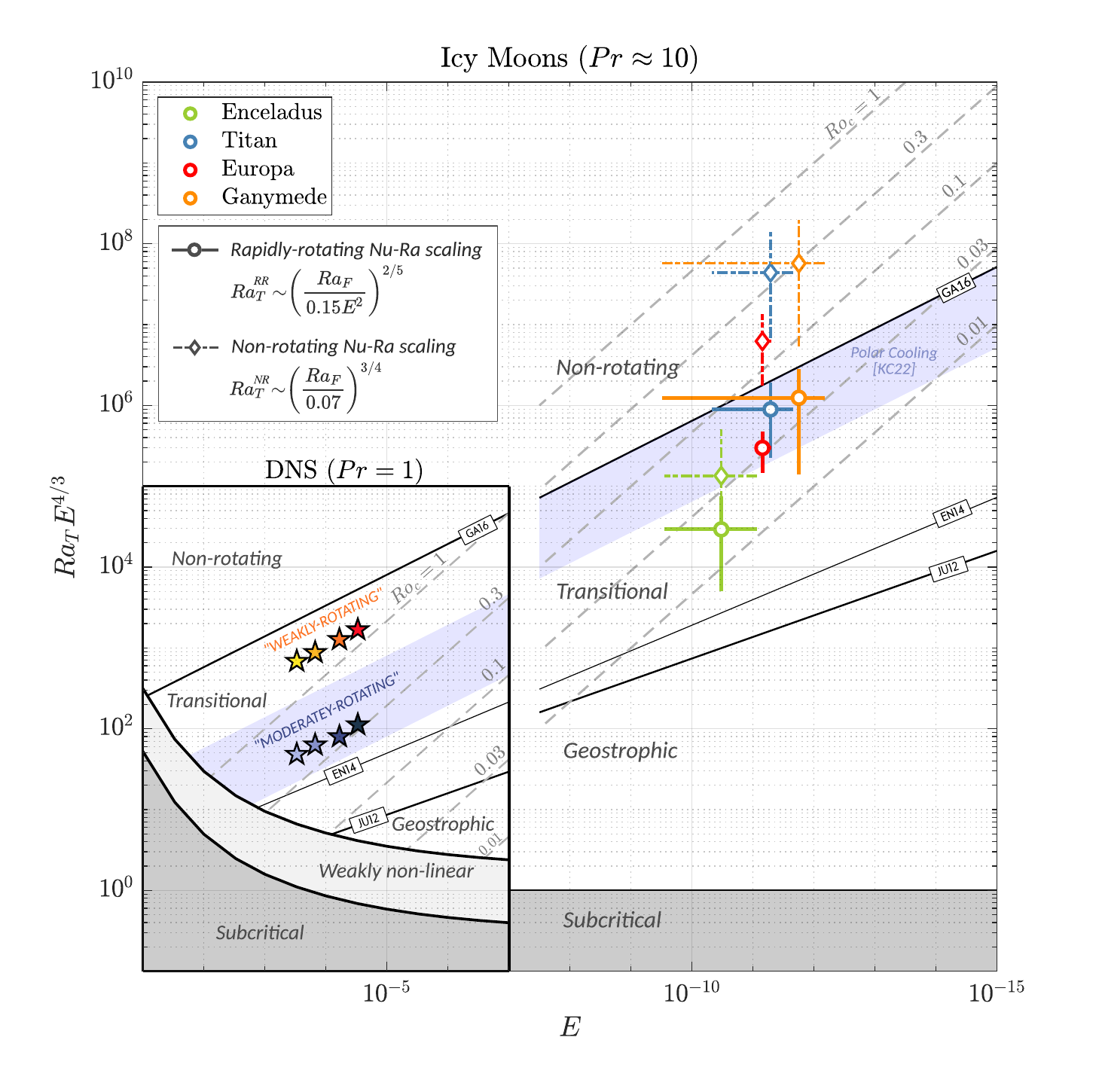}
	\caption{Regime diagram adapted from \citeA{gastine_scaling_2016} representing the supercriticality, $Ra_T E^{3/4}$, as a function of the Ekman number, for a fixed Prandtl number $Pr=1$ (Direct Numerical Simulations) and $Pr=10$ (icy moons). The grey shaded area represents a region where the flow is stable, $Ra_T<Ra_c$ (stability threshold determined by \citeA{gastine_scaling_2016} in the absence of horizontal forcing variations and for a shell aspect ratio of 0.6). Above that threshold, the flow is convectively unstable, and becomes fully non-linear once $Ra_T \gtrsim 6 Ra_c$. Different regimes of convection are observed depending on the strength of the forcing compared to rotational effects. Each regime threshold is indicated by a black continuous line, and is taken from references GA16 \cite{gastine_scaling_2016}, KC22 \cite{kvorka_numerical_2022}, EN14 \cite{ecke_heat_2014} and JU12 \cite{julien_heat_2012}. At a fixed $E$, as $Ra_T$ is increased, the convection is more vigorous, and the influence of rotation decreases, as indicated by the \textit{Geostrophic}, \textit{Transitional} and \textit{Non-rotating} regimes. The shaded blue region represents the region where \citeA{kvorka_numerical_2022} identified a polar cooling behavior, i.e. a higher heat flux at the pole. The dashed grey lines are lines of constant convective Rossby number $Ro_c = Ra_T^{1/2} E Pr^{-1/2}$. The stars represent the simulations described in the present study (see also Supplementary Table S5).}
	\label{fig:systematic-icymoons}
\end{figure}

Using scaling laws of convection (\ref{app:moonsparam}), a temperature-based Rayleigh ($Ra_T$, equation (\ref{eq:RaT})) can be estimated from the flux Rayleigh, allowing us to locate Europa on the regime diagram of Figure \ref{fig:systematic-icymoons}. Europa’s ocean falls into the transitional regime of \citeA{gastine_scaling_2016}, and is less rotationally-constrained than moons like Enceladus \cite{soderlund_ocean_2019}. {Adapting the flux-based regime diagram of} {\citeA{long_scaling_2020}} {leads to the same conclusion (see Supplementary Figure S1), but their systematic study does not allow to delineate the boundary between transitional and non-rotating regimes. Therefore, we work with the regime diagram of} \citeA{gastine_scaling_2016}. Because planetary parameters are out-of-reach with current computational capabilities, we work at more moderate parameters, chosen such that the system is in the same convective regime as Europa. Hence, we emphasize here a weakly-rotating convective regime ($Ra_F = \num{1.26e9}$, $E=\num{3e-4}$, $Pr=1$, yellow star in Figure \ref{fig:systematic-icymoons}). Because the degree of rotational influence is debated \cite{ashkenazy_dynamic_2021,bire_exploring_2022}, we provide results for a more rotationally-constrained regime (blue stars in Figure \ref{fig:systematic-icymoons}) {in section {\ref{sec:BCandrot}} and} in Supplementary Text S5. 

Note that DNS parameters are orders of magnitude wrong compared to realistic values (see also Table S4). One way to interpret the simulations' Rayleigh, Ekman and Prandtl numbers is to convert them into dimensional parameters. One possibility is to make $\Omega$, $D$, $g$, $\alpha$, $\rho$ and $c_p$ equal to the planetary values (Table {\ref{tbl:icymoonsparams}}), leading to $\nu_{\rm DNS} = \kappa_{\rm DNS} = 77.6~\si{\meter\squared\per\second}$ and $q_{\rm DNS}=\num{6.6e4} ~\si{\watt\per\meter\squared}$. These values are unphysical, meaning that our simulations are both too strongly diffusive (viscous and thermal diffusion) and too strongly forced to overcome this extra-diffusion, similarly to all other numerical studies of icy moon ocean circulation using DNS \cite<e.g., >{soderlund_ocean-driven_2014,soderlund_ocean_2019,amit_cooling_2020,kvorka_numerical_2022,terra-nova_influence_2022}. Studying dependences with input parameters is hence required for any extrapolation to the real planetary system (see section {\ref{sec:extrapolation}}, where lower Ekman numbers are considered).

From negligible to dominant tidal heating in the silicate mantle, both the average heat flux, $\langle q \rangle_h$, and the relative heat flux variations, $q^*$, should increase (Figure {\ref{fig:tidalheating}}). The range of average heat fluxes implied is taken into account in Europa's vertical error bars in Figure {\ref{fig:systematic-icymoons}}, and it does not change the convective regime into which Europa falls. We hence neglect the change in the average heat flux and work at a fixed Rayleigh number for a given sweep in $q^*$. However the amount of tidal heating relative to radiogenic heating will change the extent to which the heating of the ocean is heterogeneous. To illustrate this, we perform a sweep in $q^*$, i.e. gradually increase the bottom heat flux heterogeneity while keeping the pattern the same (top row in Figure \ref{fig:maps}). Hence, the imposed heat flux at the bottom boundary is either homogeneous (representing radiogenic heating, $q^*=0$) or increasingly heterogeneous following the tidal heating pattern ($q^*=[0.46,0.87,1.30,1.72,2.16]$). Note that we chose to go beyond $q^*=1$ to have a better physical insight on the behavior of the system, and because an interesting transition was observed around $q^* \sim 1$ for Dirichlet simulations (see Supplementary Figure S5).

\subsection{Numerical Methods}

MagIC solves for the system of equations (\ref{eq:NS}-\ref{eq:mass}) in spherical coordinates, and employs a pseudo-spectral method, using Chebyshev polynomials in the radial direction and spherical harmonics in the longitudinal and latitudinal directions. The fast spherical harmonic transform library SHTns have been employed \cite{schaeffer_efficient_2013} (URL: https://bitbucket.org/nschaeff/shtns/). The equations are integrated in time with a mixed implicit/explicit time stepping scheme. Here, we employ the commonly-used CNAB2 second-order scheme which combines a Crank-Nicolson scheme for the implicit terms and a second-order Adams-Bashforth scheme for the explicit terms (non-linear terms and Coriolis force). For full details on the numerical methods, we refer the reader to MagIC documentation available online (URL: https://magic-sph.github.io/). 

MagIC is parallelized using both OpenMP (URL: http://openmp.org/wp/) and MPI (URL: http://www.open-mpi.org/) and is designed to be used on supercomputers. The simulations were ran on the \textit{Lonestar6} system at the Texas Advanced Computing Center (URL: http://www.tacc.utexas.edu). To address differences between thermal and mechanical boundary conditions, rotation regimes, and value of $q^*$, 79 numerical runs were performed in total, as listed in Supplementary Table S5. The high Ekman number simulations are typically ran during 6 days (144 hours) on 128 CPUs. {The lowest Ekman cases were typically ran during 14 days (336 hours) on 200 CPUs.}


\section{Results}

\subsection{Mean circulation}

\begin{figure}[ht]
	\centering
	\includegraphics[width=\linewidth]{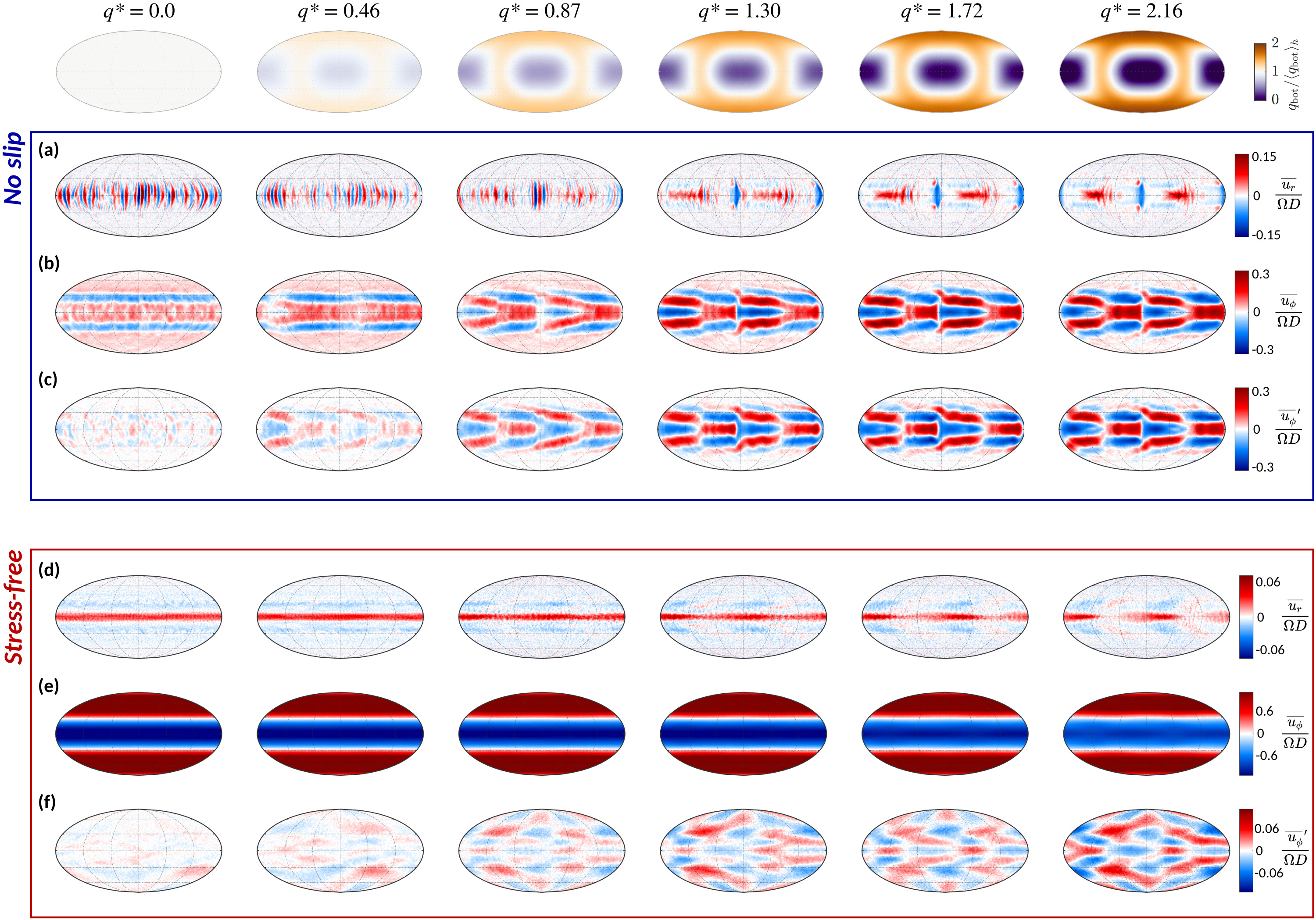}
	\caption{Mean oceanic circulation in a weakly-rotating convective regime, with no-slip and stress-free mechanical boundary conditions.  The bottom heat flux is imposed to reproduce the silicate mantle heating pattern for increasing relative amplitude of tidal heating (Figure \ref{fig:tidalheating}). All velocity maps correspond to the ocean mid-depth, and are averaged over 0.15 diffusive times to reveal the mean circulation, after a quasi steady-state has been reached. \textbf{(a,d)}: radial velocity. \textbf{(b,e)}: azimuthal velocity. \textbf{(c,f)}: azimuthal velocity from which the zonal flow was subtracted: $\overline{u_\phi}'=\overline{u_\phi} - \langle \overline{u_\phi} \rangle_\phi$. The velocity is provided in dimensionless form as global Rossby number ($D\sim 111$ km is the ocean thickness and $\Omega=\num{2.1e-5}~\si{\per\second}$ the rotation rate). Note the different color scales for the azimuthal and radial flow, and for the two boundary conditions. Extrapolation of the typical convective flow speed to Europa's parameters is discussed in Section {\ref{sec:flowspeed}}}
	\label{fig:maps}
\end{figure}

Figure \ref{fig:maps} shows the mean oceanic circulation obtained for no-slip and stress-free boundary conditions for an increasing heterogeneity $q^*$.
In no-slip simulations with homogeneous heating ($q^*=0$), moderate zonal flows are present. By ``zonal flows'', we refer to the azimuthally-averaged azimuthal velocity, $\langle u_\phi \rangle_\phi$. The flow is prograde (eastward) at the equator, retrograde (westward) at intermediate latitudes, and then prograde again in the polar regions. Axial convection with columns aligned along the rotation axis develop in the equatorial region. 

As $q^*$ is increased, the zonal flow amplitude decreases and the circulation becomes driven by the lateral variations of heating at the seafloor, a circulation called ``thermal winds solution'' hereafter, by analogy with thermal winds in atmospheres \cite{vallis_atmospheric_2017}. Thermal winds arise from the balance between buoyancy and Coriolis forces, and are described mechanistically in detail in  \citeA{dietrich_core_2016} in a configuration close to the present one. {Neglecting the time derivative (steady flow), inertia, and viscous dissipation, the curl of the Navier-Stokes equation {\eqref{eq:NS}} gives the vorticity balance}
\begin{equation}
	-\frac{2}{E} \bm{\nabla} \times \left( \bm{e}_z \times \bm{u} \right) + \frac{Ra_F}{Pr~r_0} \bm{\nabla} \times \left(T\bm{r}\right) = 0.
\end{equation}
{As detailed in} {\citeA{dietrich_core_2016}}, {taking the vertical ($z$) component of this relation and averaging along the vertical direction gives}
\begin{equation}
	\frac{2}{E} \biggl\langle \frac{\partial u_z}{\partial z} \biggr\rangle_z + \frac{Ra_F}{Pr~r_0} \bigl\langle \bm{e}_z \bm{\cdot} \left(  \bm{\nabla} \times \left(T\bm{r}\right) \right) \bigr\rangle_z = 0,
\end{equation}
{where $\langle \cdot \rangle_z = \frac{1}{2H} \int_{-H}^{+H} \cdot dz$ is the vertical averaging operator, with $H=\sqrt{r_o^2 - s^2}$ the height of the spherical shell outer boundary measured from the equatorial plane outside the tangent cylinder (the vertical cylinder tangent to the seafloor), and $s$ the cylindrical radius. The vertical velocity at $z=\pm H$ is found using non-penetration of the sloping boundaries, $u_z(+H)=-u_z(-H)=-\langle u_s \rangle_z s/H$. Evaluating the previous equation on the equatorial plane (where $s=r$) leads to}
\begin{equation}
	\langle u_s \rangle_z = -\frac{Ra_F E}{Pr} \left(\frac{r_o^2 -s^2}{2r_0s}\right) \biggl\langle \frac{\partial T}{\partial \phi} \biggr\rangle_z
\end{equation}
\cite{dietrich_core_2016}. {This equation shows that,} at the equator, in thermal wind balance, temperature gradients in longitude result in radial flows. {At the equator, the tidal heating pattern is dominated by a $m=2$ mode, with two zones of positive azimuthal heat flux gradient, and two zones of negative azimuthal gradient.} The thermal wind balance hence predicts that two zones of downwelling and two regions of upwelling {will} develop, {a feature which is indeed observed in our simulations (Figure {\ref{fig:maps}}(a) and Figure S4).} Nonlinearities due to temperature advection or inertia complicate this simple picture, and are responsible for the asymmetry between the narrow downwelling regions and the wide upwelling (see also Text S3).

With stress-free boundary conditions, the absence of viscous dissipation at the boundaries allows for the development of very strong zonal flows. For homogeneous heating ($q^*=0$), we retrieve the results of  \citeA{soderlund_ocean-driven_2014} despite their use of a different thermal boundary condition: the zonal flows are retrograde at the equator, prograde at high latitudes, and there is a mean upwelling at the equator. When $q^*$ is increased, the mean circulation is modulated by the inhomogeneous heating, but its principal features remain unchanged. We note that when we substract the zonal flow, the mean azimuthal circulation exhibits a pattern reminiscent of the thermal winds, but their amplitude is an order of magnitude smaller than the zonal flows amplitude. We call this solution the ``zonal winds solution'' hereafter.

\subsection{Vertical transfer of the bottom anomaly}

\begin{figure}[h]
	\centering
	\includegraphics[width=1\linewidth]{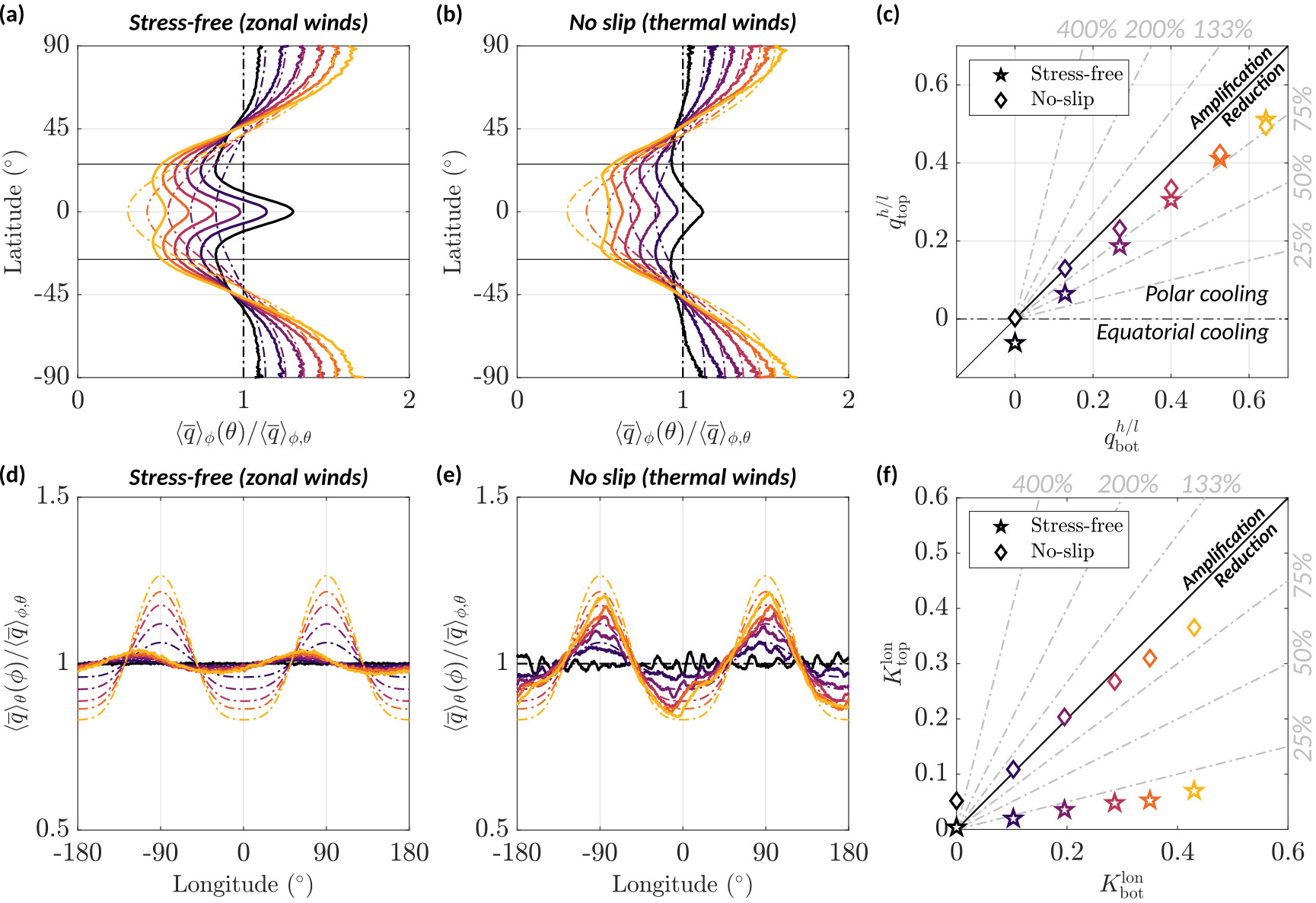}
	\caption{Relative heat flux variations measured at the top and at the bottom of the ocean for the zonal winds (first column) and the thermal winds (second column) solutions. \textbf{(a,b)} Azimuthally-averaged heat flux as a function of latitude. The dashed lines show the basal heat flux whereas the full lines are the top heat flux. From black to yellow, $q^*$ is increased from 0 to 2.16 (see Figure \ref{fig:maps}). The two black lines show the tangent cylinder latitude ($\approx 26^\circ$). \textbf{(d,e)} Same data but for the latitudinally-averaged heat flux as a function of longitude. \textbf{(c)} Latitudinal cooling parameter (equation \ref{eq:latcooling}) at the top versus at the bottom of the ocean. When $q^{h/l}<0$, the ocean is in an equatorial cooling configuration (higher heat flux at the equator). If the points lie around the 1:1 line, it means that all of the basal heat flux anomaly in latitude is retrieved at the top, and that the ocean efficiently transfers heat heterogeneities upwards. The colors have the same meaning as in other panels. \textbf{(f)} Longitudinal heat flux anomalies (equation \ref{eq:longano}) at the top versus at the bottom of the ocean.}
	\label{fig:fluxprofiles}
\end{figure}

We now investigate whether the ocean effectively transfers the tidal heating pattern from the seafloor to the ice-ocean boundary. 
Figure \ref{fig:fluxprofiles}(a,b) shows time and zonally-averaged heat flux profiles at the bottom and top boundaries. In the zonal winds solution, with homogeneous basal heating, the heat flux peaks at the equator due to the equatorial upwelling, consistently with  \citeA{soderlund_ocean-driven_2014}. As $q^*$ is increased, even if an equatorial peak persists, the top heat flux follows the basal heat flux and becomes greater at the pole even for the smallest heterogeneity considered ($q^*= 0.43$). The thermal winds solution exhibits a very similar behavior.

To quantify the latitudinal cooling configuration, we define the dimensionless parameter
\begin{linenomath*}
	\begin{equation}
		q^{h/l} = \frac{\langle \overline{q} \rangle_{\rm high} - \langle \overline{q} \rangle_{\rm low}}{\langle \overline{q} \rangle_{\rm high} + \langle \overline{q} \rangle_{\rm low}},
		\label{eq:latcooling}
	\end{equation}
\end{linenomath*}
where $\overline{q}$ is the time-averaged heat flux at the top or at the bottom of the ocean, $\langle \cdot \rangle_{\rm low}$ is the average over latitudes lower than $10^\circ$, and $\langle \cdot \rangle_{\rm high}$ over latitudes greater than $80^\circ$. When $q^{h/l}<0$ the heat flux is larger in the equatorial region, and the ocean is said to be in an equatorial cooling configuration. Conversely, the ocean is in polar cooling if $q^{h/l}>0$.  Figure \ref{fig:fluxprofiles}(c) confirms that the ocean is in an equatorial cooling configuration only for the homogeneous heating case. Furthermore, the ocean efficiently transfers the latitudinal variations of basal heating up to the ice-ocean boundary, with $q^{h/l}_{\rm top}\approx 0.75 ~q^{h/l}_{\rm bot}$. For dominant tidal heating, the heat flux at the poles would be systematically higher than at the equator, i.e. the ocean would be in a \textit{polar cooling} configuration.

We next measure the relative amplitude of longitudinal heat flux variations at the bottom and at the top of the ocean:
\begin{linenomath*}
	\begin{equation}
		K_j^{lon} = \vert \langle \overline{q}_j \rangle_{\rm lat} ^\textup{max} - \langle \overline{q}_j \rangle _{\rm lat} ^\textup{min} \vert /\langle \overline{q}_j \rangle_h,~~~\textup{with}~~ j \in \{\textup{top,bot}\}
		\label{eq:longano}
	\end{equation}
\end{linenomath*}
Here, $\langle \cdot \rangle_\textup{lat}$ denotes average in latitude and $\langle \cdot \rangle_h$ denotes horizontal average on the sphere. Figure \ref{fig:fluxprofiles}(d-f) show that in the thermal winds solution, longitudinal variations are almost entirely transferred upward thanks to the mean equatorial loops (Supplementary Text S3). In the zonal winds solution, longitudinal variations are smeared, and only about 20\% of the original anomaly persists at the ice-ocean boundary. 

\subsection{Sensitivity to boundary conditions and rotation regime}

\label{sec:BCandrot}

In each convective regime, we investigated the sensitivity to the thermal boundary conditions {(BC)} by imposing the temperature {(Dirichlet BC)} at the seafloor instead of the heat flux. We show in the Supplementary Information (Text S4) that the results are very similar. The only notable exception is that in the weakly-rotating regime, with stress-free, temperature-imposed boundary conditions, an abrupt transition occurs between the zonal and thermal winds solutions (Figure \ref{fig:systematic}(c) and Supplementary Figure S5). The origin of this transition will be investigated further, but is beyond the scope of the present study as flux-imposed boundary conditions are the most planetary relevant. 

{Given the uncertainty in Europa's ocean convective regime} ({\ref{app:moonsparam}}), {there has been a debate on its cooling configuration (polar versus equatorial). We focused here on a single, weakly-rotating regime, but from Figure {\ref{fig:systematic-icymoons}}, Europa falls within the polar cooling regime identified by } \citeA{kvorka_numerical_2022}. {To sample this polar cooling regime, we performed more rotationally-constrained simulations, while remaining in the transitional regime of}  \citeA{gastine_scaling_2016}. {The regime of these simulations is called ``moderately-rotating'' hereafter and correspond to blue stars on Figure {\ref{fig:systematic-icymoons}}. We show in the Supplementary Information (Text S5) that the results are qualitatively unaffected by the change in the convective regime: thermal winds develop with no-slip boundary conditions, whereas zonal flows are dominant with stress-free boundary conditions. Latitudinal heat flux variations from the seafloor are mirrored at the ice-ocean boundary, and longitudinal variations are reduced if zonal winds develop.}

\subsection{Extrapolation towards planetary regimes}
\label{sec:extrapolation}

\subsubsection{Heat anomaly}

{To explore the effect of going towards more planetary relevant parameters (smaller Ekman and larger Rayleigh), we reduce $E$ and increase $Ra_T$ while remaining in a given convective regime, i.e. keeping the same degree of rotational influence (Figure {\ref{fig:systematic-icymoons}}). For both the weakly-rotating (yellow-red stars) and moderately-rotating (blue stars) regimes, we explored four different Ekman numbers $E \in [{\num{3e-4}}, {\num{1.5e-4}},{\num{6e-5}},{\num{3e-5}}]$. To keep computational costs reasonable, this systematic study was done for Dirichlet, stress-free boundary conditions only. We chose Dirichlet thermal BCs because the choice of $Ra_T$ to be in the correct convective regime can be directly picked from the regime diagram of GA16 (Figure {\ref{fig:systematic-icymoons}}), and stress-free BCs because they allow us to sample both the zonal winds and thermal winds solutions at the same time given the existence of the transition. The corresponding set of simulations is listed in Supplementary Table S5 and represented in Figure {\ref{fig:systematic-icymoons}}. As an equivalent to $q^*$ for Neumann simulations, we define the bottom temperature anomaly $\Delta T^* = {\Delta T_h}/{\Delta T_v}$ where ${\Delta T_h}$ is the peak-to-peak temperature difference at the seafloor. For each pair $(Ra_T,E)$, we perform an entire sweep in $\Delta T^*$ from 0 to 1.63 (see Supplementary Text S4.1), except for the smallest Ekman number for which we select a few $\Delta T^*$.}

\begin{figure}[h!]
	\centering
	\includegraphics[width=\linewidth]{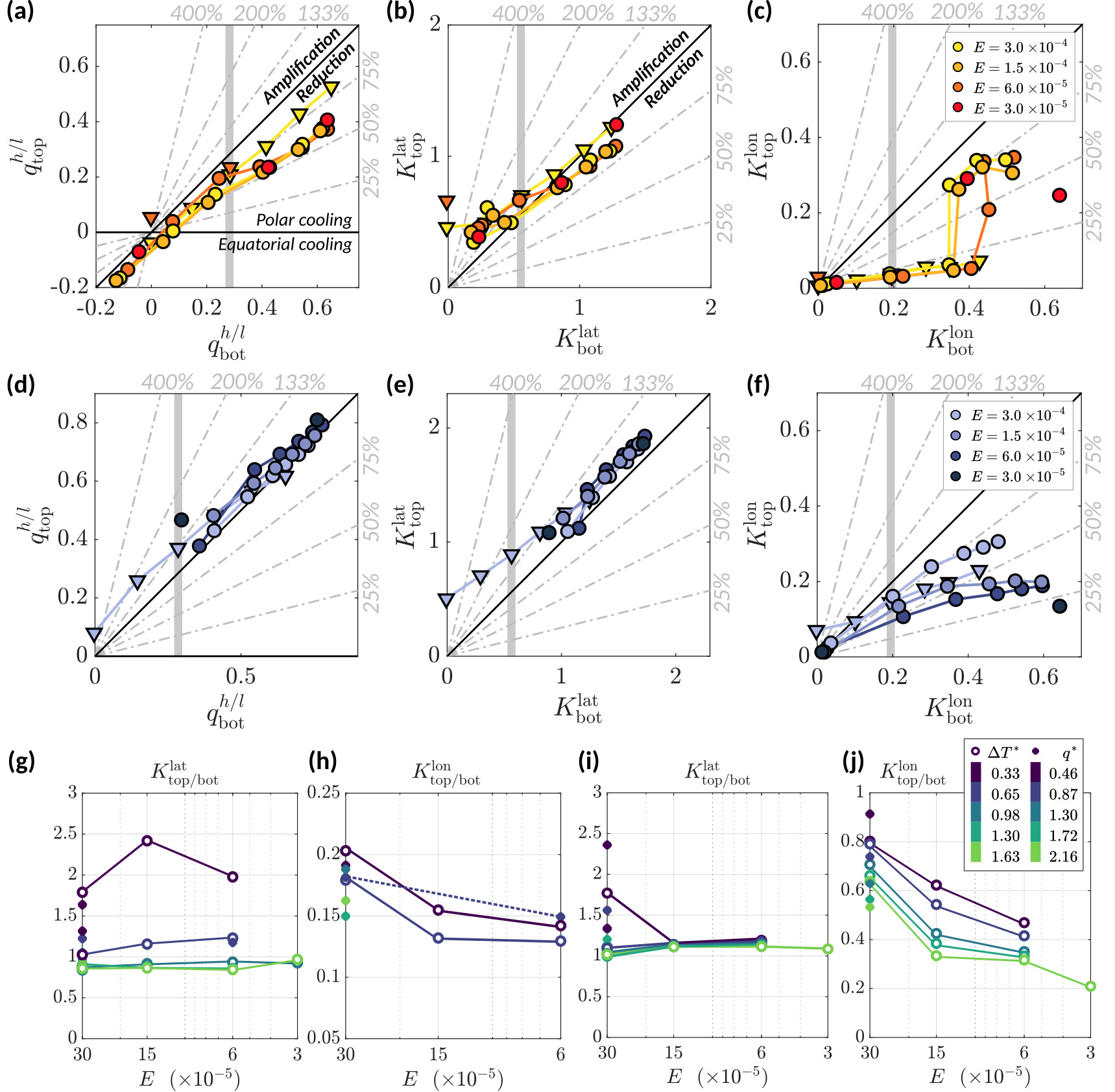}
	\caption{ {Systematic for stress-free boundary conditions (BCs) in weakly-rotating (a-c) and moderately-rotating (d-f) regimes. Circles: Dirichlet bottom BC, Triangles: Neumann bottom BC. Colors correspond to the colors of the DNS stars in Figure {\ref{fig:systematic-icymoons}}. \textbf{(a,d)} $q^{h/l}$ at the top and at the bottom of the ocean.  \textbf{(b,e)} Anomaly in latitude at the top and at the bottom of the ocean. $K^{\rm lat}$ is defined as $K^{\rm lon}$ (equation {\ref{eq:longano}}), except with a longitudinal average instead of the latitudinal average.  \textbf{(c,f)} Anomaly in longitude at the top and at the bottom of the ocean. In panel (c), the jump in the longitudinal anomaly at the top of the ocean is due to a transition from the zonal winds to the thermal winds solution. The transition occurs for all Ekman numbers investigated. In all panels, the vertical grey line represent bottom anomalies for pure tidal heating in the silicate mantle. \textbf{(g-j)} Fraction of the anomaly retrieved at the top of the ocean for simulations with the zonal winds solution. Open symbols: Dirichlet bottom boundary condition. Filled  symbols and dashed line: Neumann bottom boundary condition. \textbf{(g)} Weakly-rotating regime, latitudinal anomaly. \textbf{(h)} Weakly-rotating regime, longitudinal anomaly. \textbf{(i)} Moderately-rotating regime, latitudinal anomaly. \textbf{(j)} Moderately-rotating regime, longitudinal anomaly.}}
	\label{fig:systematic}
\end{figure}

Figure {\ref{fig:systematic}}(a-f) represents the relative heat flux anomalies at the top versus at the bottom of the ocean for all the simulations of the systematic study (circles), plus the simulations with Neumann BCs described previously (triangles). Figure {\ref{fig:systematic}}(g-j) shows the fraction of the anomaly retrieved at the top of the ocean, as a function of the Ekman number. Figure {\ref{fig:systematic}}(a,b,d,e,g,i) shows that in both the weakly-rotating and moderately-rotating regimes, the latitudinal anomaly is efficiently transferred across all Ekman numbers investigated. It is slightly reduced for the weakly-rotating cases because of the equatorial cooling configuration (higher heat flux at the equator) of this regime when the bottom heating is homogeneous \cite{kvorka_numerical_2022}. The heat flux latitudinal anomaly is on the contrary slightly amplified at the top of ocean for moderately-rotating cases due to the polar cooling configuration of this regime when the bottom heating is homogeneous \cite{kvorka_numerical_2022}. 

The preservation of latitudinal heating variations up to the ice-ocean interface is qualitatively in line with \citeA{gastine2023latitudinal} who underline that different convective dynamics occur in the equatorial region compared to the polar regions, even with a homogeneous bottom forcing. This regionalization of convection is proposed to be due to the increased misalignment between the rotation vector and gravity from the pole to the equator, as well as the extent of the tangent cylinder, fixed by the shell aspect ratio. The more decoupled the polar and equatorial regions are, the better they should reflect their respective (local) bottom heating, and the better large-scale latitudinal variations should be preserved. To verify this hypothesis, our study calls for more detailed analysis of the heat transport (de)coupling between polar and equatorial regions in weakly-rotating, moderately-rotating and rapidly-rotating regimes. Besides, to ascertain extrapolation of this behavior to real Europa parameters ($E\sim \num{e-11}$, $Ra_F\sim\num{e27}$), accurate asymptotic trends have to be established on a larger range than what we could cover here ($E \in [{\num{3e-4}},{\num{3e-5}}]$), which is a computational challenge.

{For longitudinal variations, in the weakly-rotating regime, Figure {\ref{fig:systematic}}(c) shows that it is hard to argue for a clear change in the efficiency of the longitudinal variations transfer as Ekman is reduced. This is especially true for the thermal winds solution (after the jump in panel (c)), for which the fraction of the longitudinal anomaly retrieved at the top of the ocean, $K_{\rm top/bot}^{\rm lon} = K_{\rm top}^{\rm lon}/K_{\rm bot}^{\rm lon}$, stays close to about 75\%. For the zonal winds solution however, $K_{\rm top/bot}^{\rm lon}$ decreases as we explore more planetary relevant regimes. In the weakly-rotating regime and for the cases representing pure tidal heating (grey vertical line on Figure {\ref{fig:systematic}}(c)), $K_{\rm top/bot}^{\rm lon}$  decreases from 20\% to 15.5\% to 14.2\% at $E=[\num{3e-4},\num{1.5e-4},\num{6e-5}]$ repectively (Figure {\ref{fig:systematic}}(h)). In the moderately-rotating regime, $K_{\rm top/bot}^{\rm lon}$ decreases from about 80\% to 47\% from $E=\num{3e-4}$ to $\num{6e-5}$ (Figure {\ref{fig:systematic}}(j)).}

\begin{figure}[ht]
	\centering
	\includegraphics[width=\linewidth]{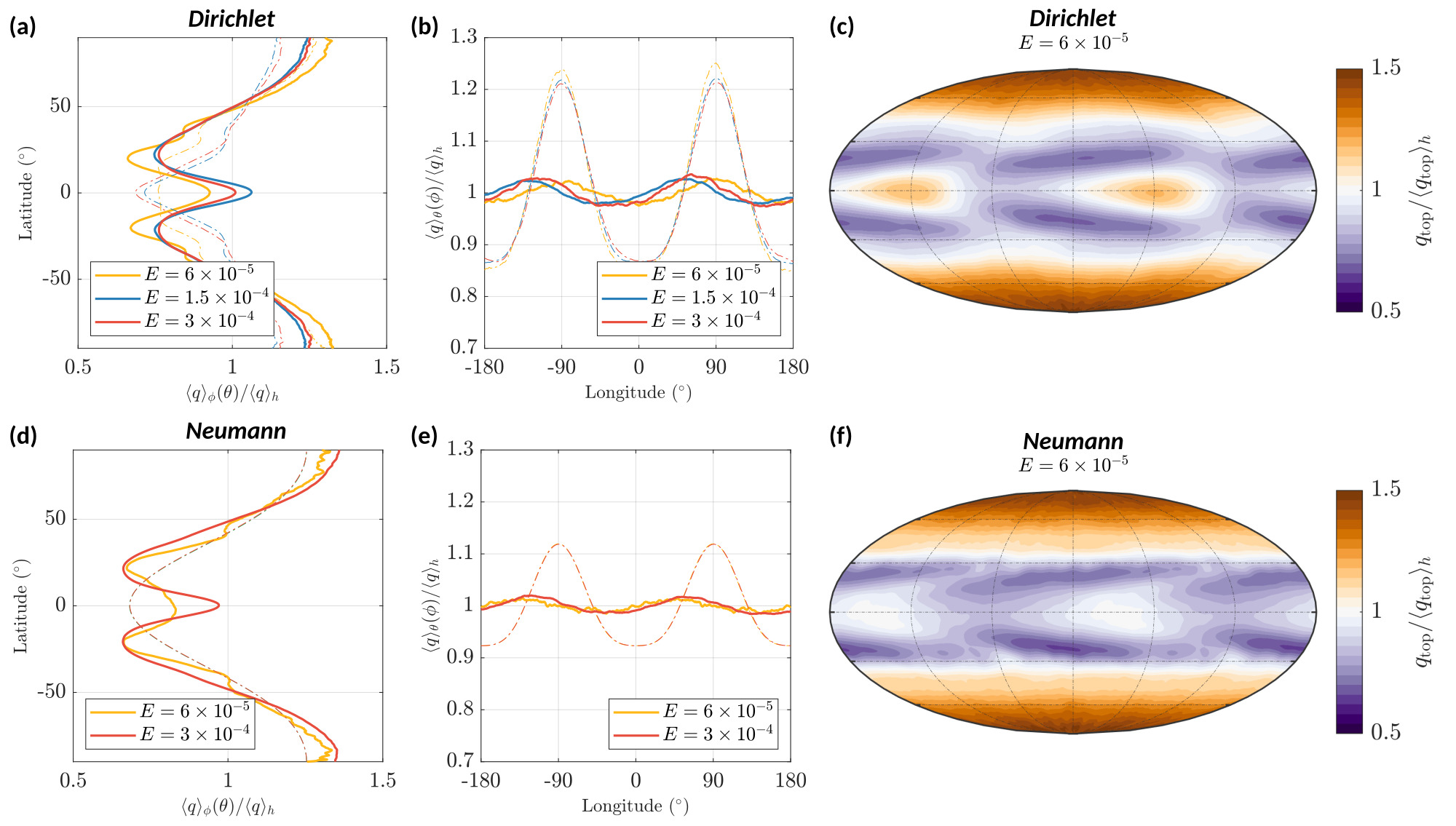}
	\caption{Heat flux at the top of the ocean for simulations with stress-free boundary conditions and a tidally-dominant scenario ($\Delta T^*=0.65$ and $q^*=0.87$). \textbf{Top row}: Dirichlet bottom boundary condition. \textbf{Bottom row}: Neumann bottom boundary condition. \textbf{(a,d)} Latitudinal heat flux profiles at the bottom (dashed lines) and at the top (full lines) of the ocean. $\langle \cdot \rangle_\phi$ denotes average in longitude and $\langle \cdot \rangle_h$ denotes horizontal average on the sphere. \textbf{(b,e)} Longitudinal heat flux profiles at the bottom (dashed lines) and at the top (full lines) of the ocean. $\langle \cdot \rangle_\theta$ denotes average in latitude.  \textbf{(c,f)} Normalized heat flux maps at the top of the ocean.}
	\label{fig:smallEkman}
\end{figure}

{Figure {\ref{fig:smallEkman}} shows heat flux profiles and maps for weakly-rotating cases in the pure tidal heating scenario, for Dirichlet and Neumann BCs ($\Delta T^*=0.65$ and $q^*=0.87$), at various Ekman numbers.  As already said above, the latitudinal anomaly is transferred entirely up to the ice-ocean boundary for all Ekman numbers investigated (Figure {\ref{fig:smallEkman}}(a,d)), and the ocean is in a polar cooling configuration. Figure {\ref{fig:smallEkman}}(b,c,e,f) also shows that significant longitudinal variations are present at the top of the ocean, even if $K_{\rm top/bot}^{\rm lon}$ decreases. For Dirichlet BCs, $K_{\rm top/bot}^{\rm lon}$  decreases from 18\% to 13.0\% at $E=[\num{3e-4},\num{6e-5}]$ repectively. With Neumann BCs, it decreases from 18\% at $E=\num{3e-4}$ to 15\% at $E=\num{6e-5}$ (see also Figure  {\ref{fig:systematic}}(h)). Because of their computational cost,  we did not perform additional cases at even smaller Ekman and higher Rayleigh. It remains to be investigated if the longitudinal anomaly reaches a plateau at even smaller Ekman numbers and higher Rayleigh, or if the heat flux at the top of the ocean becomes zonally-symmetric.}

\subsubsection{Flow speed}
\label{sec:flowspeed}

Typical convective flow speed in planetary regimes can be estimated using scaling laws for the convective Reynolds number, defined here as a dimensionless root-mean-squared (rms) velocity: $Re_c = \sqrt{2E_k/V_s}=u_{\rm rms}D/\nu$, where $E_k$ is the time-averaged, dimensionless kinetic energy from which the contribution of the axisymmetric mode is subtracted, and $V_s$ is the dimensionless fluid volume. In the transitional regime between rapidly-rotating and non-rotating convection, there is no clear asymptotic scaling for the convective Reynolds number {\cite{gastine_scaling_2016}}. {$Re_c$ should nevertheless be bounded by the non-rotating and rapidly rotating predictions, $Re_c^{nr}\sim Ra_T^{1/2}$ and  $Re_c^{rr} \sim (Ra_FE^{1/2}Pr^{-2})^{2/5}$, respectively} {\cite{gastine_scaling_2016}}. {Our parameter space exploration shows a convective Reynolds number scaling $Re_c^{\rm fit} \sim 0.65(Ra_F E^{1/2})^{0.45}$, corresponding to intermediate values between $Re_c^{rr}$ and $Re_c^{nr}$ (Supplementary Figure S10). Note that our fit does not include any dependence on $Pr$ because we did not vary this parameter. Using the parameters for Europa listed in Table {\ref{tbl:icymoonsparams}}, the predicted convective Reynolds are  $Re_c^{nr}\sim\num{3.1e10}-\num{1.3e11}$, $Re_c^{rr}\sim \num{4.2e7}-\num{2.2e8}$ and $Re_c^{\rm fit}\sim \num{2.1e9}-\num{1.3e10}$ (see also Supplementary Table S4). This would correspond to a range of flow speeds from $u_{\rm rms}^{rr}\sim$ 0.7 to 3.5 mm/s, to $u_{\rm rms}^{nr}\sim$ 0.5 to 2 m/s. Using our scaling, we find speeds from $u_{\rm rms}^{\rm fit}\sim$ 3.5 cm/s to 22 cm/s.} \citeA{jansen_energetic_2023-1} found, based on energetic constraints, that ocean currents in Europa should be of the order of a few centimeters per second, i.e., intermediate values between our scaling and the rapidly-rotating scaling.

\subsection{Ice thickness}
\label{sec:icethickness}

To illustrate the possible impact on ice shell equilibrium, two end-member scenarios are considered: 1) a radiogenic-dominated scenario ($q^*=0$) where tidal heating in the silicate mantle is negligible and 2) a tidally-dominated scenario ($q^*= 0.87$), where tidal heating significantly exceeds radiogenic heating. Oceanic heat flux maps used as inputs for each end-member scenario are represented in Figure \ref{fig:fluxice}(a,d), where we focus on the zonal winds solution {represented in Figure {\ref{fig:maps}} (weakly-rotating, Neumann, stress-free cases $q^*=0$ and $q^*=0.87$). We showed that relative latitudinal heat flux variations obtained from the simulations are robust for the range of parameters explored. Assuming that this pole-to-equator heat flux difference continues to hold for real Europa parameters (see section {\ref{sec:extrapolation}}), we compute oceanic heat flux maps by extracting the relative heat flux variations from the simulations, and multiplying them by realistic average heat fluxes for each end-member. We caution again that longitudinal variations could be erased once in planetary regimes and should hence be viewed as upper bounds. In scenario 1 (2), we assume 5 mW/m$^2$ (37 mW/m$^2$) as the average oceanic heat flux at the ice-ocean interface (see also \ref{app:iceinputs}). 

{The resulting ice thickness is computed using the model of} \citeA{nimmo_global_2007}, {described in the Supplementary Text S2. In a nutshell, this model assumes a purely conductive ice shell, where the local ice thickness depends on the surface temperature (which varies with latitude), the temperature at the bottom of the shell, assumed to be at the melting temperature, the oceanic heat flux, and the volumetric tidal heat production in the ice. It neglects altogether heat diffusion in the horizontal (latitudinal and longitudinal directions), viscous relaxation, variations of the melting temperature with pressure, and latent heat due to phase change. The laterally varying fields used as inputs for the ice thickness model -- oceanic heat flux, tidal strain rate and surface temperature -- are represented in Figure {\ref{fig:iceinputs}} in} {\ref{app:iceinputs}}.

\begin{figure}[h]
	\centering
	\includegraphics[width=1\linewidth]{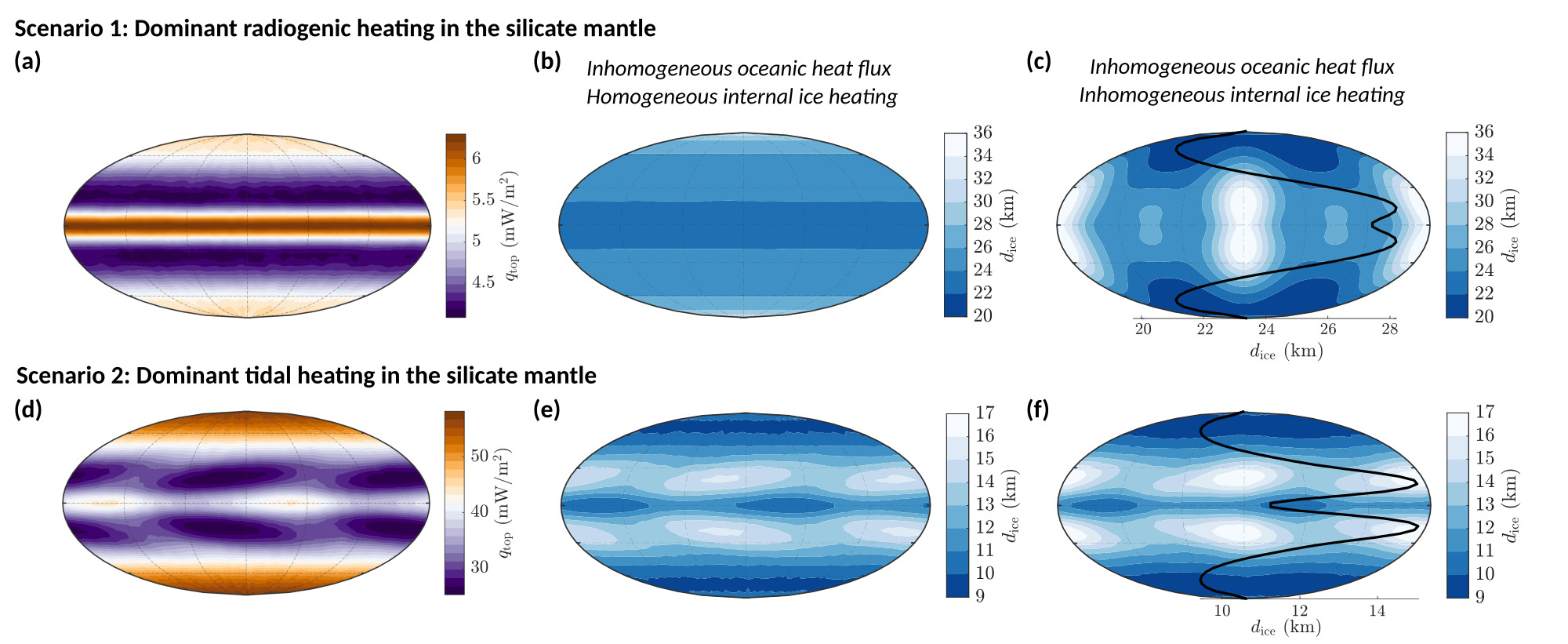}
	\caption{Heat flux at the ice-ocean boundary in the zonal winds solution and corresponding ice thickness assuming a purely radially conducting ice shell. \textbf{(a)} Time-averaged heat flux at the top of the ocean for $q^*=0$ (pure radiogenic heating scenario) in the zonal winds solution (stress-free). The maps are averaged over 0.15 diffusive times, as in Figure \ref{fig:maps}. \textbf{(b)} Corresponding ice thickness maps assuming that the internal heating in the ice is laterally homogeneous (see Figure \ref{fig:iceinputs}). \textbf{(c)} Ice thickness maps taking into account that the internal heating in the ice is laterally inhomogeneous. \textbf{(d,e,f)} Similar maps for the case $q^*=0.87$ (dominant tidal heating scenario). The black curves on panels (c) and (f) represent the zonally-averaged ice thickness, to emphasize latitudinal variations. Given the idealized nature of our model, we warn the reader that neglected processes within the ice shell (e.g., viscous flow, convection) as well as feedback effects of phase changes on the ocean dynamics could act to smooth the thickness variations obtained here.}
	\label{fig:fluxice}
\end{figure}

Both oceanic heating and internal ice heating vary laterally and could induce ice thickness variations. To estimate their relative contributions, we first focus on the ice thickness resulting from a heterogeneous oceanic heating, while the internal ice heating is taken homogeneous and equal to its average value (Figures \ref{fig:fluxice}(b,e)). In the scenario where radiogenic heating is dominant, the ice shell is thinner in the equatorial region due to higher oceanic heat flux and higher surface temperature. If tidal heating from the mantle is considered, this latitudinal difference is reduced or even reversed and longitudinal variations appear. Unlike latitudinal variations, longitudinal heat flux variations are sensitive to the choice of mechanical boundary conditions. If strong zonal winds develop, they could be swept out once planetary regimes are reached (Section {\ref{sec:extrapolation}}). 

Endogenic processes within the ice will add complexity to this picture. To illustrate this, we add lateral variations of internal ice heating, arising from the spatially variable tidal strain rate (Figure \ref{fig:iceinputs}(a)). Figure \ref{fig:fluxice}(c) shows that the dominant radiogenic heating scenario leads to thick ice, between 21 and 37 km (average thickness of 25 km). The {total} internal ice heating is {$5.6\times 10^{11}$ W (equivalent to 19 mW/m$^2$ at the ice-ocean boundary), and is} responsible for almost all of the ice thickness variations, which thus follow the tidal strain rate pattern within the ice. The fact that the oceanic heat flux is higher by a few \si{\milli\watt\per\meter\squared} at the equator leads, nevertheless, to slightly thinner ice at the equator.  The dominant tidal heating scenario  (Figure \ref{fig:fluxice}(f)) would lead to globally thinner ice, because of the higher oceanic heat flux, with an average ice shell thickness of 12 km and variations between 9 and 17 km. {The total internal ice heating is $1.6\times 10^{11}$ W (5 mW/m$^2$ at the ice-ocean boundary).} The ice thickness variations are almost entirely determined by the oceanic heat flux. Minimum thicknesses are found at high latitudes and at the equator and largest thicknesses at low latitudes ($\sim 25^\circ$), at the longitudes of the sub and anti-jovian points. The tidal heating within ice slightly reinforces the variations already induced by the oceanic heat flux.  The thermal winds solution leads to similar conclusions ({Supplementary Figure S3}). {As discussed in section {\ref{sec:extrapolation}}, longitudinal variations obtained in Figure {\ref{fig:fluxice}}(c,f) represent upper bounds as they may not be maintained in planetary regimes.} 

{The ice thickness maps of Figure {\ref{fig:fluxice}} are used to illustrate which heating variations dominate the thermal equilibrium of a conductive ice shell (oceanic heating versus tidal heating in the ice). Again, we warn the reader that they should not be considered as an accurate quantitative prediction of ice thickness variations on Europa. In the real system, many other processes within the ice shell as well as feedback effects on the ocean dynamics will likely act to smooth the thickness variations obtained here. Regarding endogenic ice processes, lateral thickness variations could be erased by viscous flow} \cite{nimmo_non-newtonian_2004,kamata2017interior}, {or by solid-state convection} \cite<e.g., >{mckinnon_convective_1999,tobie_tidally_2003,ashkenazy_dynamics_2018}. {From Galileo limb profiles,} \citeA{nimmo_global_2007} {argue that lateral shell thickness variations do not exceed 7 km on Europa. This is consistent with the idea that the thickness variations obtained here are upper bounds of what could be expected on Europa.}

{As discussed in} \citeA{kamata2017interior}, {melting and freezing of ice need to occur to actively maintain any ice thickness gradient against the lateral ice flow.} {Equation (6) in } \citeA{nimmo_non-newtonian_2004} {provides an estimate of the ice flow timescale assuming a Newtonian rheology:} 
\begin{equation}
	\tau = \frac{\eta_b}{\Delta \rho g \delta^3 k^2},~~~\delta=\frac{R T_m D}{E \ln(T_m/T_s)},~~~k=\frac{2\pi}{\lambda},
\end{equation}
{where $\eta_b$ is the basal viscosity of ice, $\Delta \rho$ is the density contrast between the ice shell and the ocean, $R$ is the universal gas constant, $T_m$ is the melting temperature, $T_s$ is the surface temperature, $D$ is the average ice thickness, $E$ is the activation energy and $\lambda$ is the wavelength of the topography.}
{Using the same parameters as in the ice thickness model (Supplementary Table S2), $\Delta \rho = 100$ kg/m$^3$, and $\lambda = 2450$ km (pole-to-equator topography), we obtain $\tau_{\rm flow}=$ 115 Myr for a 12km-thick ice, and 13 Myr for a 25km-thick ice. For a steady-state to exist, the melting/freezing timescale needs to be equal to the ice flow timescale} \cite{kamata2017interior,shibley2023infer}. {The heat flux at the ice-ocean boundary that would provide energy to sustain this melt rate can be estimated as $q_{\rm melt} \sim \rho_i L \Delta H/\tau_{\rm flow}$, where $\rho_i= 920$ kg m$^{-3}$ is density of ice, $L=330$ kJ kg$^{-1}$ is latent heat of ice and $\Delta H$ is the ice thickness contrast. Assuming a typical pole-to-equator change of ice thickness $\Delta H\sim5$ km, we obtain $q_{\rm melt}\sim 0.4$ mW m$^{-2}$ for a 12km-thick ice, and $q_{\rm melt}\sim 3.8$ mW m$^{-2}$ for a 25km-thick ice, which is not unrealistic.} {That being said, given the huge uncertainty in the basal ice viscosity, the ice flow could be an order of magnitude faster or slower} \cite{nimmo_non-newtonian_2004}. {Hence, a 5 km pole-to-equator thickness contrast could either be impossible to sustain due to a fast viscous relaxation, or be effectively sustained by corresponding melting/freezing at the ice-ocean boundary. Note that the Newtonian rheology underestimates the flow timescale compared to more realistic rheologies, especially for long wavelength topographies} \cite{nimmo_non-newtonian_2004}, {supporting further the possibility of a steady-state.} 

{Regarding feedback effects on the ocean dynamics, an important missing process in our model is the phase change at the ice-ocean boundary. The freezing point depression with pressure, and therefore with ice thickness, could induce a feedback on the underlying convection. Ice thickness variations of 5 km on Europa would lead to lateral temperature variations of about 460 mK at the ice-ocean boundary (equation (5) in} \citeA{kang_how_2022}). {These lateral temperature variations could force horizontal convection in the ocean, in a similar fashion as lateral heat flux variations at the seafloor. As investigated by } \citeA{kang2023modulation}, {the formation of warm water under thin ice regions, and cold water under thick ice regions, may build up a stratified layer underneath the ice and prevent the bottom heat flux pattern from reaching the ice-ocean boundary} \cite<see also>{zhu_influence_2017}. As discussed above, any ice thickness variation would need to be actively maintained by freezing and melting to compensate the viscous flow, and the associated brine and fresh-water injections could then drive its own large-scale circulation \cite{lobo_pole--equator_2021,kang_how_2022}. Including salinity effects also adds the possibility of double-diffusive convection \cite{vance_layering_2005,wong_layering_2022}. The large-scale topography due to ice thickness variations could also induce its own feedback on the ocean dynamics. Finally, mechanically-driven and electro-magnetically-driven flows could also arise, but how they would compete with buoyancy-driven flows remains an open problem \cite<>[and references therein]{soderlund_physical_2024}.

\section{Conclusion}

In this study, we investigate if rotating thermal convection in Europa's ocean can transfer the heterogeneous tidal heating pattern at the seafloor up to the ice-ocean boundary. We use an idealized model where we neglect salinity and the possible feedback of phase change and topography at the ice-ocean boundary. Under these assumptions, we found that latitudinal heat flux variations from the seafloor are essentially mirrored at the ice-ocean boundary. Hence, if tidal heating in the silicate mantle is dominant compared to radiogenic heating, it would impose a polar cooling configuration for Europa {(higher oceanic heat flux at the poles)}, independently of the convective regime considered. Unlike radiogenic heating, tidal heating in the silicate mantle of Europa could also participate in producing longitudinal heat flux variations. The amplitude of these longitudinal variations is however strongly dependent on the ability of thermal winds to develop in planetary regimes, {and they could be smoothed, particularly in the presence of zonal winds. Quantifying if longitudinal variations persist in more extreme regimes than those investigated here} remains a computational challenge.

{Using the conductive ice thickness model of} \citeA{nimmo_global_2007}, we conclude with two end-member scenarios. If radiogenic heating is dominant in the silicate mantle, the heat flux variations from the ocean would be too weak to influence the thermal equilibrium of the ice shell, which would be mostly controlled by internal tidal heating in the ice. If tidal heating in the silicate mantle is in excess of the radiogenic contribution, the oceanic heat flux would be high enough to control the ice thickness. The ocean heat flux would at least reflect latitudinal variations of heat flux at the seafloor, leading to {relatively} thin polar ice.

As detailed in the previous section, the present ocean model is missing important physical processes. {Coupled models able to take into account simultaneously the phase change and the nonlinear ocean dynamics driven by various forcing mechanisms are sorely needed to investigate feedback effects. The freezing point depression, but also the large-scale topography at the ice-ocean boundary should be incorporated in future models of buoyancy-driven ocean circulation. Furthermore, tides, libration and precession can excite a variety of waves in the ocean} \cite<see the review by>[and references therein]{soderlund_physical_2024}. {The nonlinear interaction of these waves can trigger boundary or bulk-filling turbulence, and induce dissipation which could contribute to the heat budget of the ocean. The feedback between convection and mechanically-driven flows deserves further attention.}

Studying {coupled} ice and ocean processes constitute exciting challenges for planetary modellers, both from the fundamental perspective and to prepare for interpretation of \textit{Europa Clipper} \cite{phillips_europa_2014,roberts2023exploring} and JUICE \cite{grasset_jupiter_2013} observations. Future observations from \textit{Europa Clipper} may be able to disentangle between the two scenarios proposed here, 1) radiogenic-dominated heating in the silicate mantle, resulting in a thick ice shell, with a higher equatorial oceanic heat flux and 2) tidally-dominated heating in the silicate mantle, resulting in a thin ice shell, with a higher polar oceanic heat flux. This might be achieved by using motional magnetic induction \cite{vance2021magnetic}, constraining the depth of the ice-ocean interface and thermal variations in the ice using the ice-penetrating radar \cite{kalousova_2017_radar,blankenship_reason_2018}, constraining the ice shell thickness using gravity measurements \cite{mazarico_2023_europa},  or estimating the asymmetry of the ocean thickness using the induced magnetic field \cite{styczinski_perturbation_2022}.

\appendix

\section{Icy Moon parameters and regimes of rotating convection}
\label{app:moonsparam}

Table \ref{tbl:icymoonsparams} contains the dimensional and dimensionless physical parameters used to plot the regime diagram of Figure \ref{fig:systematic-icymoons}. The range of Europa's plausible heat fluxes (6 to 46 \si{\milli\watt\per\meter\squared}) is determined the following way: 	
Assuming that the radiogenic heating in Europa's mantle is of 6-7 \si{\milli\watt\per\meter\squared} at the seafloor \cite{tobie_tidally_2003}, the lower bound corresponds to the case of a negligible tidal heating in the silicate mantle. The upper bound corresponds to the upper bound of the log-normal distribution considered by \citeA{howell_likely_2021}, i.e. a tidal heat flux of about $10^{-1.5}\approx 32$ \si{\milli\watt\per\meter\squared} at the base of the ice (see their Figure 3). At the seafloor, this would lead to a heat flux of 39 \si{\milli\watt\per\meter\squared}, to which a radiogenic heating of 7 \si{\milli\watt\per\meter\squared} is added. This upper bound accounts for the possibility of Io-like tidal dissipation in Europa's mantle \cite{howell_likely_2021}, i.e., a total dissipated power of 1000 GW. The other parameters are taken from \citeA{soderlund_ocean_2019}. For Europa, with the average parameters and their bounds listed in Table \ref{tbl:icymoonsparams}, we obtain the dimensionless parameters $Pr\approx 11$, $E= \num{7e-12}$ ($\in[\num{5.0e-12},\num{9.1e-12}]$) and $Ra_F=\num{5.4e27}$ ($\in[\num{6.3e26},\num{2.8e28}]$).

\begin{table}[h]
	\centering
	\resizebox{\textwidth}{!}{%
		\begin{tabular}{lcccc}
			\multicolumn{2}{l}{\textbf{Physical parameters}}  \\
			& Enceladus & Titan & Europa & Ganymede \\
			\toprule[0.5pt] 
			$g~(\si{\meter\per\second\squared})$ & 0.1 & 1.4 & 1.3 & 1.4 \\
			$\Omega~(\num{e-5}\si{\per\second})$ & 5.3 & 0.46 & 2.1 & 1.0 \\
			$\nu~(\num{e-6}\si{\meter\squared\per\second})^{(a)}$ & 1.8 & 1.8 & 1.8 & 1.8 \\
			$\kappa~(\num{e-7}\si{\meter\squared\per\second})^{(a)}$ & 1.4 & 1.8 & 1.6 & 1.8 \\
			$R~(\si{\kilo\meter})$ & 252 & 2575 & 1561 & 2631 \\
			$D_{\rm ice}~(\si{\kilo\meter})$ & $30.5\in[10,51]$ & $99.5\in[50,149]$ & $17.5\in[5,30]$ & $69.5\in[5,157]$ \\
			$D_{\rm ocean}~(\si{\kilo\meter})$ & $32\in[11,63]$ & $275\in[91,420]$ & $111\in[97,131]$ & $319\in[24,518]$
			\\
			$q~(\si{\milli\watt\per\meter\squared})$ & $48.5\in[16,83]$ & $17\in[14,20]$ & $26\in[6,46]$  & $61\in[15,107]$ \\
			$\rho~(\si{\kilogram\per\meter\cubed})$ & $1000\in[1000,1110]$ & $1125\in[1110,1240]$ & $1040\in[1040,1150]$ & $1125\in[1110,1190]$ \\
			$c_p~(\num{e3}\si{\joule\per\kilogram\per\kelvin})$ & $4.2\in[3.6,4.2]$ & $3.3\in[2.1,3.6]$ & $3.9\in[3.3,3.9]$ & $3.4\in[2.1,3.7]$ \\
			$\alpha~(\num{e-4} \si{\per\kelvin})$ & $0.6\in[0.1,1.3]$ & $2.0\in[0.4,4.2]$ & $2.0\in[1.9,2.5]$ & $3.0\in[1.9,4.4]$ \\		
			\toprule					 								
			\multicolumn{2}{l}{\textbf{Dimensionless parameters}} \\
			& Enceladus & Titan & Europa & Ganymede \\
			\toprule[0.5pt]
			$\eta = 1-\frac{D_{\rm ocean}}{R-D_{\rm ice}}$ & $0.86\in[0.74,0.95]$ & $0.89\in[0.83,0.96]$ & $0.93\in[0.92,0.94]$ & $0.87\in[0.80,0.99]$ \\
			$Pr = \nu/\kappa$ & $13$ & $10$ & $11$ & $10$ \\
			$E = \frac{\nu}{\Omega D_{\rm ocean}^2}$ & $\num{3.3e-11}$ & $\num{5.2e-12}$ & $\num{7.0e-12}$ & $\num{1.8e-12}$ \\	
			& 	$\in[\num{8.6e-12},\num{2.8e-10}]$ & $\in[\num{2.2e-12},\num{4.7e-11}]$ & $\in[\num{5.0e-12},\num{9.1e-12}]$ & $\in[\num{6.7e-13},\num{3.1e-10}]$ \\
			$Ra_F = \frac{\alpha g q D_{\rm ocean}^4}{\rho c_p \nu \kappa^2}$ & $\num{2.1e24}$ & $\num{1.3e29}$ & $\num{5.4e27}$ & $\num{1.2e30}$ \\	
			& 	$\in[\num{1.4e21},\num{1.3e26}]$ & $\in[\num{2.1e26},\num{2.7e30}]$ & $\in[\num{6.3e26},\num{2.8e28}]$ & $\in[\num{5.2e24},\num{3.5e31}]$ \\
			$Ra_T^{NR} = \left( \frac{Ra_F}{0.07} \right)^{3/4}$ & $\num{1.3e19}$ & $\num{4.9e22}$ & $\num{4.6e21}$ & $\num{2.7e23}$ \\	
			& 	$\in[\num{5.4e16},\num{2.9e20}]$ & $\in[\num{4.0e20},\num{4.9e23}]$ & $\in[\num{9.3e20},\num{1.6e22}]$ & $\in[\num{2.5e19},\num{3.4e24}]$ \\	
			$Ra_T^{RR} = \left( \frac{Ra_F}{0.15 E^2} \right)^{2/5}$ & $\num{2.7e18}$ & $\num{1.0e21}$ & $\num{2.2e20}$ & $\num{5.8e21}$ \\
			& 	$\in[\num{2.7e16},\num{4.3e19}]$ & $\in[\num{1.3e19},\num{6.7e21}]$ & $\in[\num{7.6e19},\num{5.6e20}]$ & $\in[\num{6.6e17},\num{4.8e22}]$ \\	
			$Ra_T^{KC22} = b  \frac{Ra_F^\beta E^{3\beta-2}}{Pr^{2\beta-1}} $ & $\num{2.1e18}$ & $\num{1.6e21}$ & $\num{2.6e20}$ & $\num{8.0e21}$ \\	
			& 	$\in[\num{1.9e16},\num{3.4e19}]$ & $\in[\num{2.1e19},\num{1.1e22}]$ & $\in[\num{7.6e19},\num{7.1e20}]$ & $\in[\num{1.4e18},\num{7.1e22}]$ \\			
			\bottomrule
			
			\multicolumn{5}{l}{$^{(a)}${Dimensionless parameters are based on molecular values of viscosity and diffusivity. }} \\
			\multicolumn{5}{l}{In global ciculation models (GCM) where small-scale turbulence is unresolved,} \\
			\multicolumn{5}{l}{its effect is parametrized by using effective viscosities and diffusivities which are orders of} \\
			\multicolumn{5}{l}{ {magnitude larger than molecular values, and often different in horizontal and vertical directions} \cite<e.g.,>[]{kang2023modulation}.} \\
			\end{tabular}
	}
	\caption{\label{tbl:icymoonsparams} Dimensional and dimensionless physical parameters used to plot the regime diagram of Figure \ref{fig:systematic-icymoons}. The dimensional physical parameters are taken from \citeA{soderlund_ocean_2019} except for Europa's heat flux: $g$ is the gravitational acceleration, $\Omega$ the rotation rate, $\nu$ the kinematic viscosity, $\kappa$ the thermal diffusivity, $R$ the moon's radius, $D_{ice}$ the ice thickness, $D_{\rm ocean}$ the ocean thickness, $q$ the heat flux per unit area, $\rho$ the ocean density, $c_p$ the ocean's heat capacity, and $\alpha$ the ocean's thermal expansivity. For the dimensionless parameters, $\eta$ is the aspect ratio, $Pr$ the Prandlt number, $E$ the Ekman number, $Ra_F$ the flux Rayleigh number. $Ra_T^{NR}$, $Ra_T^{RR}$ and $Ra_T^{KC22}$ are estimates of temperature Rayleigh numbers from the flux Rayleigh number, as defined in \ref{app:moonsparam}. For $Ra_T^{KC22}$, we use the values of $b$ and $\beta$ given in \citeA{kvorka_numerical_2022}: $b\approx 14.73$ and $\beta \approx 0.525$.}
\end{table}

Systematic studies of rotating thermal convection in spherical shells have shown that various regimes of convection can take place, depending on the degree of rotational influence. Figure \ref{fig:systematic-icymoons} reproduces the regime boundaries summarized by \citeA{gastine_scaling_2016} (GA16 hereafter). These boundaries were identified using Dirichlet boundary conditions, therefore the relevant Rayleigh number is a temperature one,
\begin{linenomath*}
	\begin{equation}
		Ra_T = \frac{\alpha g_o   \Delta T_v  D^3 }{\nu \kappa} = \frac{Ra_F}{Nu},
		\label{eq:RaT}
	\end{equation}
\end{linenomath*}
where $\Delta T_v = T_{\rm bot} - T_{\rm top}$ is the imposed vertical temperature difference, and $Nu$ is the Nusselt number
\begin{linenomath*}
	\begin{equation}
		Nu = \frac{qD}{k \Delta T_v}.
	\end{equation}
\end{linenomath*}
$k=\rho c_p \kappa$ is the thermal conductivity of the fluid, and $q$ is the total heat flux per unit area. In Dirichlet simulations, $\Delta T_v$ is imposed, and the total heat flux $q$ exceeds the conductive heat flux $~k \Delta T_v/D$ because of convection. Therefore, the Nusselt number is greater than one. To estimate the temperature Rayleigh number of icy moons and locate them in the regime diagram of Figure \ref{fig:systematic-icymoons}, the Nusselt number needs to be estimated, but the temperature constrast across the ocean is unknown. To overcome this, and following  \citeA{soderlund_ocean_2019}, we use Nusselt-Rayleigh scaling laws determined by systematic numerical studies \cite{gastine_scaling_2016}: in the non-rotating limit, $Nu \sim 0.07 Ra_T^{1/3}$ and in the limit of rapidly-rotating convection, $Nu \sim 0.15 Ra_T^{3/2} E^2$. Substituting $Nu \sim Ra_F/Ra_T$ in these scaling laws lead to
\begin{linenomath*}
	\begin{equation}
		Ra_T^{NR} = \left(\frac{Ra_F}{0.07}\right)^{3/4}
		\label{eq:NRscaling}
	\end{equation}
\end{linenomath*}  
in the non-rotating (NR) limit, and
\begin{linenomath*}
	\begin{equation}
		Ra_T^{RR} = \left(\frac{Ra_F}{0.15 E^2}\right)^{2/5}
		\label{eq:RRscaling}		
	\end{equation}
\end{linenomath*} 
in the rapidly-rotating (RR) limit. A third estimate is proposed following \citeA{kvorka_numerical_2022}, where a diffusivity-free scaling for the convective Rossby number, $Ro_c$, is employed:
\begin{linenomath*}
	\begin{equation}
		Ro_c^2 = \frac{Ra_T E^2}{Pr} \sim b (Ra_F^*)^\beta 
	\end{equation}
\end{linenomath*} 
(their equation B.8). Here, $Ra_F^*$ is a modified flux Rayleigh number, $Ra_F^* = Ra_F E^3 Pr^{-2}$. Substituting in the previous equation, and using $b=14.73$ and $\beta =0.525$ \cite{kvorka_numerical_2022}, we obtain a third relation,
\begin{linenomath*}
	\begin{equation}
		Ra_T^{KC22} = b Ra_F^\beta E^{3\beta -2} Pr^{1-2\beta} \approx 14.73 \frac{Ra_F^{0.525}}{E^{0.425} Pr^{-0.05}}
		\label{eq:KC22scaling}
	\end{equation}
\end{linenomath*} 
For Europa's ocean, we obtain $Ra_T^{NR} \approx \num{4.6e21}$, $Ra_T^{RR} \approx \num{2.2e20}$, $Ra_T^{KC22} \approx \num{2.6e20}$ (Table \ref{tbl:icymoonsparams}). Using these values, we represent Europa in the regime diagram of Figure \ref{fig:systematic-icymoons}. Since the last two estimates are in fact close, we only represent $Ra_T^{NR}$ and $Ra_T^{RR}$. A similar procedure can be applied to other icy moons, they are represented in the diagram for completeness even if they are not the focus of the present study. Corresponding physical and dimensionless parameters are listed in Table \ref{tbl:icymoonsparams}. Europa possibly falls within the non-rotating or upper transitional regime of \citeA{gastine_scaling_2016}. The simulations presented in the main text are chosen in the transitional regime (yellow star in the DNS insert on Figure \ref{fig:systematic-icymoons}). At an Ekman number of $E=\num{3e-4}$ and $Pr=1$, this corresponds to $Ra_T = \num{3.4e7}$. To choose the corresponding flux Rayleigh, instead of using the previous scaling laws, we use the actual Nusselt number measured at the end of the corresponding Dirichlet simulation ($Nu \approx 37$), leading to $Ra_F = Ra_T Nu = \num{1.26e9}$. Note that using the scaling laws (\ref{eq:NRscaling}-\ref{eq:KC22scaling}), we would have obtained $Ra_F \in [\num{7.71e8},\num{9.10e10}]$, consistent with the fact that we are in a regime intermediate between non-rotating and rapidly-rotating. More detailed discussion about rotating convection regimes is available in the Supplementary Information, and results are provided for a more rotationally-constrained possibility there as well (Supplementary Text S5).

\section{Ice thickness model inputs}
\label{app:iceinputs}

Figure \ref{fig:iceinputs} represents maps of the physical quantities used as inputs for the ice thickness model of \citeA{nimmo_global_2007}, i.e. the surface temperature, the tidal strain rate and the oceanic heat flux. {The ice thickness model is described in the Supplementary Information, Text S2.}

As mentioned in the main text, because the extreme parameters of icy moon oceans are out of reach of Direct Numerical Simulations (DNS) with current computational capabilities, our DNS are too strongly diffusive (viscous and thermal diffusion) and too strongly forced.
Since the dimensional heat flux imposed in the DNS is unphysically high, using the DNS heat flux in combination with realistic ice parameters is inconsistent and would lead to unphysically thin ice. In the ice thickness model, we therefore use the DNS \textit{relative} heat flux variations, multiplied with a realistic mean heat flux (see Figures \ref{fig:iceinputs}(e,f)) to obtain realistic dimensional ice thicknesses. This procedure implicitly assumes that the relative variations will stay the same when reaching planetary regimes (smaller Ekman, higher Rayleigh).  Section \ref{sec:extrapolation} shows that for the range of parameters studied here, this is true for latitudinal variations, but longitudinal variations could be smoothed. The longitudinal variations presented here should hence be considered as upper bounds.

In the radiogenic-dominated scenario, an average heat flux of $5 ~\si{\milli\watt\per\meter\squared}$ at the top of the ocean is used. The corresponding heat flux maps are represented on Figure \ref{fig:iceinputs}(c,e). At the top of the ocean, the heat flux would globally vary between 4 and 7 mW/m$^2$, with a maximum at the equator, and no longitudinal variations. In the tidally-dominated scenario, we use the upper bound of 46 \si{\milli\watt\per\meter\squared} at the bottom of the ocean (Table \ref{tbl:icymoonsparams}), i.e. 37 mW/m$^2$ at the top of the ocean.

\begin{figure}[h!]
	\centering
	\includegraphics[width=\linewidth]{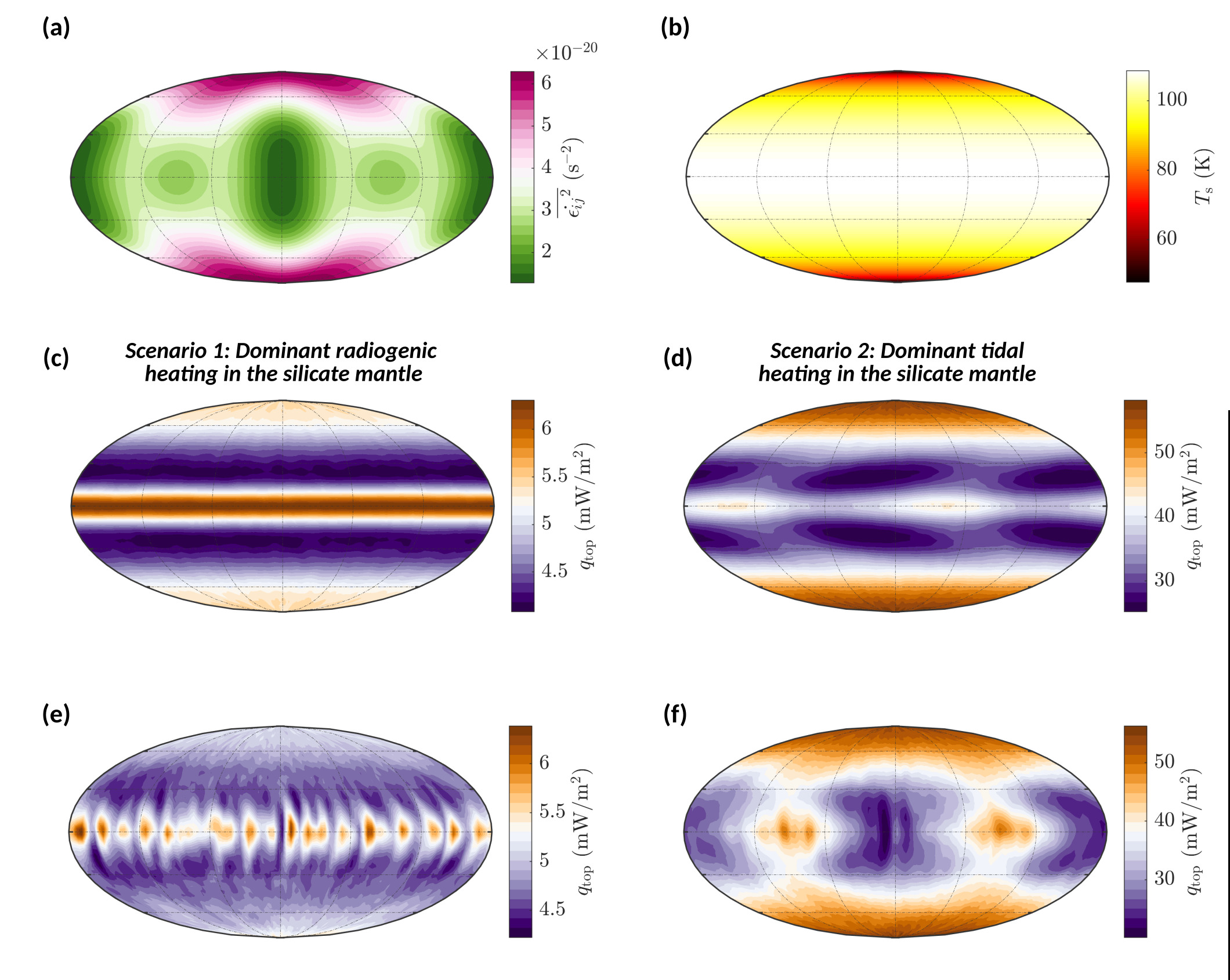}
	\caption{Inputs of the ice thickness model. \textbf{(a)} Ice tidal strain rate calculated from \citeA{ojakangas_thermal_1989}. {Average value $\overline{\epsilon_{ij}^2}=3.3 \times 10^{-20}$ s$^{-2}$. This value is used for the cases of Figure {\ref{fig:fluxice}} where homogeneous internal ice heating is assumed.} \textbf{(b)} Temperature at the surface of the ice calculated from \citeA{ojakangas_thermal_1989}. \textbf{(c,e)} Oceanic heat flux map estimated from the relative heat flux variations of the DNS with $q^*=0$ (pure radiogenic heating) and stress-free (c) and no-slip (e) boundary conditions. \textbf{(d,f)} Oceanic heat flux estimated from simulations with $q^*=0.87$ (dominant tidal heating) and stress-free (d) and no-slip (f) boundary conditions.}
	\label{fig:iceinputs}
\end{figure}

%
%

\section*{Open Research Section}
The numerical code used to perform the ocean convection simulations is an open-source software called MagIC \cite{wicht_inner-core_2002,christensen_numerical_2001,schaeffer_efficient_2013}. It is developed on GitHub (URL: https://magic-sph.github.io/) and can be used, modified and redistributed under the terms of the GNU GPL v3 licence. MagIC is available in this in-text data citation reference: \citeA{MagIC6.0_zenodo}. Datasets for this research are available in this in-text data citation reference: \citeA{lemasquerier_zenodo}.

\section*{Conflict of Interest Statement}
The authors have no conflicts of interest to declare.

\acknowledgments
The authors thank the University of Texas Institute of Geophysics and Jackson School of Geosciences for funding. The authors acknowledge the Texas Advanced Computing Center (TACC) at The University of Texas at Austin for providing HPC and  visualization resources that have contributed to the research results reported within this paper (URL: http://www.tacc.utexas.edu). We thank Dunyu Liu for his help in installing MagIC on TACC. The authors acknowledge ideas and advice from the participants in the Next-Generation Planetary Geodesy workshop organized by the W.M. Keck Institute for Space Studies, which facilitated this collaboration. The authors thank Francis Nimmo (editor) and three anonymous reviewers whose comments and suggestions helped improve and clarify this manuscript.

%
%

\nocite{andrade_viscous_1910,aurnou_connections_2020,castillo-rogez_tidal_2011,efroimsky_tidal_2012,jackson_shear_2004,jackson_grain-size-sensitive_2002,king_boundary_2009,renaud_increased_2018,sundberg_composite_2010,tan_high-temperature_2001}

\bibliography{IcyMoons}



\section*{Supplementary material}

| Parameter | Unit | Value (average if varying) |
|---|---|---|
| $\mu$ | GPa | 4 |
| $\overline{\dot{\epsilon}_{ij}}(\theta,\phi)$ | $s^{-1}$ | $2 \times 10^{-10}$, Appendix B in Ojakangas and Stevenson (1989) |
| $\omega$ | $s^{-1}$ | $2.048 \times 10^{-5}$ |
| $\eta_b$ | $Pa\,s$ | $3 \times 10^{14}$ |
| $E$ | $J\,mol^{-1}$ | $59 \times 10^3$ |
| $R$ | $J\,K^{-1}\,mol^{-1}$ | $8.314$ |
| $T_s(\theta)$ | K | 100, Equations (A.17) and (A.18) in Ojakangas and Stevenson (1989) |
| $T_m$ | K | 270 |
| $k_{ice}$ | $W\,m^{-1}\,K^{-1}$ | 3 |
| $q_{oc}$ | $W\,m^{-2}$ | $5 \times 10^{-3}$ [radiogenic end-member] |
| | | $37 \times 10^{-3}$ [tidal end-member] |

**Table S3:** Parameters of the ice thickness model (taken from Nimmo et al. (2007)). Maps of the input fields for the ice thickness model are provided in Appendix B in the main text.

| | Europa | DNS WR | DNS MR |
|---|---|---|---|
| $\eta$ | $0.92 - 0.94$ | 0.9 | 0.9 |
| $P_r$ | 11 | 1 | 1 |
| $E$ | $5.0 \times 10^{-12} - 9.1 \times 10^{-12}$ | $3 \times 10^{-4} - 3 \times 10^{-5}$ | $3 \times 10^{-4} - 3 \times 10^{-5}$ |
| $Ra_F$ | $6.3 \times 10^{26} - 2.8 \times 10^{28}$ | $1.3 \times 10^9 - 2.4 \times 10^{11}$ | $1.8 \times 10^7 - 2.4 \times 10^9$ |
| $Ra_T$ | $7.6 \times 10^{19} - 1.6 \times 10^{22}$ | $3.4 \times 10^7 - 1.8 \times 10^9$ | $2.4 \times 10^6 - 1.2 \times 10^8$ |
| $Ro_c = Ra_T^{1/2} E Pr^{-1/2}$ | $1.9 - 27 \times 10^{-2}$ | $1.3 - 1.7$ | $0.33 - 0.46$ |
| $Ro_\lambda = Ra_T E^{8/5} Pr^{-3/5}$ | $26 - 5400$ | $79 - 100$ | $5.5 - 7.0$ |
| $Ro_F^{NR} = (Ra_F E^3 Pr^{-2})^{1/3}$ | $8.7 - 55 \times 10^{-4}$ | $0.19 - 0.32$ | $4.0 \times 10^{-2} - 7.8 \times 10^{-2}$ |
| $Ro_F^{RR} = (Ra_F E^3 Pr^{-2})^{1/5}$ | $1.5 - 4.4 \times 10^{-2}$ | $0.37 - 0.51$ | $0.15 - 0.22$ |
| $Ro^* = (Ra_F E^3 Pr^{-2}/8)^{1/2}$ | $9.0 \times 10^{-6} - 1.5 \times 10^{-4}$ | $6.5 \times 10^{-2} - 8.0 \times 10^{-2}$ | $7.8 \times 10^{-3} - 8.1 \times 10^{-3}$ |
| $Re_c^{nr} \sim Ra_T^{1/2}$ | $3.1 \times 10^{10} - 1.3 \times 10^{11}$ | $5.8 \times 10^3 - 4.2 \times 10^4$ | $1.5 \times 10^3 - 1.1 \times 10^4$ |
| $Re_c^{rr} \sim (Ra_F E^{1/2} Pr^{-2})^{2/5}$ | $4.2 \times 10^7 - 2.2 \times 10^8$ | $860 - 4400$ | $160 - 700$ |
| $Re_c^{fit} \sim 0.65(Ra_F E^{1/2})^{0.45}$ | $2.1 \times 10^9 - 1.3 \times 10^{10}$ | $1300 - 8200$ | $190 - 1000$ |

**Table S4:** Dimensionless parameters for the DNS in comparison to estimated parameters for Europa. "WR" refers to weakly-rotating (yellow-red stars on Figure 2), "MR" refers to moderately-rotating (blue stars on Figure 2). The various Rossby numbers are defined in Supplementary Text S5. Dimensionless parameters involved in the definitions are listed in Table A1 of the main text for Europa and Table S5 for the DNS.

**Table S5:** Parameters of the 79 numerical simulations (file uploaded separately). For all of the simulations, $Pr = 1$ and $\eta = 0.9$. TBC refers to "thermal boundary condition" for the bottom boundary (the temperature is imposed at the top boundary in all runs). N means Neumann (in which case $Ra_F$ and $q^*$ are given), and D means Dirichlet (in which case $Ra_T$ and $\Delta T^*$ are given). MBC refers to "mechanical boundary condition", for the top and bottom boundaries. SF means stress-free, and NS means no-slip. $N_r$ is the number of radial grid points, $N_\phi$ is the number of azimuthal grid points, the number of grid points in latitude being $N_\theta = N_\phi/2$. These simulations are represented on Figure 2 in the main text.

| Label | $E$ | $Ra_T$ or $Ra_F$ | TBC | MBC | $N_r$ | $N_\phi$ | Regime | $\Delta T^*$ or $q^*$ |
|---|---|---|---|---|---|---|---|---|
| N1 | $3 \times 10^{-4}$ | $1.26 \times 10^9$ | N | SF | 97 | 640 | Weakly-rotating | 0.0 |
| N2 | $3 \times 10^{-4}$ | $1.26 \times 10^9$ | N | SF | 97 | 640 | Weakly-rotating | 0.46 |
| N3 | $3 \times 10^{-4}$ | $1.26 \times 10^9$ | N | SF | 97 | 640 | Weakly-rotating | 0.87 |
| N4 | $3 \times 10^{-4}$ | $1.26 \times 10^9$ | N | SF | 97 | 640 | Weakly-rotating | 1.30 |
| N5 | $3 \times 10^{-4}$ | $1.26 \times 10^9$ | N | SF | 97 | 640 | Weakly-rotating | 1.72 |
| N6 | $3 \times 10^{-4}$ | $1.26 \times 10^9$ | N | SF | 97 | 640 | Weakly-rotating | 2.16 |
| N7 | $3 \times 10^{-4}$ | $1.26 \times 10^9$ | N | NS | 97 | 1024 | Weakly-rotating | 0.0 |
| N8 | $3 \times 10^{-4}$ | $1.26 \times 10^9$ | N | NS | 97 | 1024 | Weakly-rotating | 0.46 |
| N9 | $3 \times 10^{-4}$ | $1.26 \times 10^9$ | N | NS | 97 | 1024 | Weakly-rotating | 0.87 |
| N10 | $3 \times 10^{-4}$ | $1.26 \times 10^9$ | N | NS | 97 | 1024 | Weakly-rotating | 1.30 |
| N11 | $3 \times 10^{-4}$ | $1.26 \times 10^9$ | N | NS | 97 | 1024 | Weakly-rotating | 1.72 |
| N12 | $3 \times 10^{-4}$ | $1.26 \times 10^9$ | N | NS | 97 | 1024 | Weakly-rotating | 2.16 |
| N13 | $3 \times 10^{-4}$ | $1.8 \times 10^7$ | N | SF | 97 | 1280 | Moderately-rotating | 0.0 |
| N14 | $3 \times 10^{-4}$ | $1.8 \times 10^7$ | N | SF | 97 | 1280 | Moderately-rotating | 0.46 |
| N15 | $3 \times 10^{-4}$ | $1.8 \times 10^7$ | N | SF | 97 | 1280 | Moderately-rotating | 0.87 |
| N16 | $3 \times 10^{-4}$ | $1.8 \times 10^7$ | N | SF | 97 | 1280 | Moderately-rotating | 1.30 |
| N17 | $3 \times 10^{-4}$ | $1.8 \times 10^7$ | N | SF | 97 | 1280 | Moderately-rotating | 1.72 |
| N18 | $3 \times 10^{-4}$ | $1.8 \times 10^7$ | N | SF | 97 | 1280 | Moderately-rotating | 2.16 |
| N19 | $3 \times 10^{-4}$ | $1.8 \times 10^7$ | N | NS | 129 | 1280 | Moderately-rotating | 0.0 |
| N20 | $3 \times 10^{-4}$ | $1.8 \times 10^7$ | N | NS | 129 | 1280 | Moderately-rotating | 0.46 |
| N21 | $3 \times 10^{-4}$ | $1.8 \times 10^7$ | N | NS | 129 | 1280 | Moderately-rotating | 0.87 |
| N22 | $3 \times 10^{-4}$ | $1.8 \times 10^7$ | N | NS | 129 | 1280 | Moderately-rotating | 1.30 |
| N23 | $3 \times 10^{-4}$ | $1.8 \times 10^7$ | N | NS | 129 | 1280 | Moderately-rotating | 1.72 |
| N24 | $3 \times 10^{-4}$ | $1.8 \times 10^7$ | N | NS | 129 | 1280 | Moderately-rotating | 2.16 |
| N25 | $6 \times 10^{-5}$ | $4.9 \times 10^{10}$ | N | SF | 201 | 2048 | Weakly-rotating | 0.0 |
| N26 | $6 \times 10^{-5}$ | $4.9 \times 10^{10}$ | N | SF | 201 | 2048 | Weakly-rotating | 0.87 |
| D1 | $3 \times 10^{-4}$ | $3.4 \times 10^7$ | D | NS | 97 | 1024 | Weakly-rotating | 0.0 |
| D2 | $3 \times 10^{-4}$ | $3.4 \times 10^7$ | D | NS | 97 | 1024 | Weakly-rotating | 0.33 |
| D3 | $3 \times 10^{-4}$ | $3.4 \times 10^7$ | D | NS | 97 | 1024 | Weakly-rotating | 0.65 |
| D4 | $3 \times 10^{-4}$ | $3.4 \times 10^7$ | D | NS | 97 | 1024 | Weakly-rotating | 0.98 |
| D5 | $3 \times 10^{-4}$ | $3.4 \times 10^7$ | D | NS | 97 | 1024 | Weakly-rotating | 1.30 |
| D6 | $3 \times 10^{-4}$ | $3.4 \times 10^7$ | D | NS | 97 | 1024 | Weakly-rotating | 1.63 |
| D7 | $3 \times 10^{-4}$ | $2.4 \times 10^6$ | D | NS | 129 | 1280 | Moderately-rotating | 0.0 |

| | | | | | | | | |
|---|---|---|---|---|---|---|---|---|
| D8 | $3 \times 10^{-4}$ | $2.4 \times 10^{6}$ | D | NS | 129 | 1280 | Moderately-rotating | 0.33 |
| D9 | $3 \times 10^{-4}$ | $2.4 \times 10^{6}$ | D | NS | 129 | 1280 | Moderately-rotating | 0.65 |
| D10 | $3 \times 10^{-4}$ | $2.4 \times 10^{6}$ | D | NS | 129 | 1280 | Moderately-rotating | 0.98 |
| D11 | $3 \times 10^{-4}$ | $2.4 \times 10^{6}$ | D | NS | 129 | 1280 | Moderately-rotating | 1.30 |
| D12 | $3 \times 10^{-4}$ | $2.4 \times 10^{6}$ | D | NS | 129 | 1280 | Moderately-rotating | 1.63 |
| D13 | $3 \times 10^{-4}$ | $3.4 \times 10^{7}$ | D | SF | 97 | 640 | Weakly-rotating | 0.0 |
| D14 | $3 \times 10^{-4}$ | $3.4 \times 10^{7}$ | D | SF | 97 | 640 | Weakly-rotating | 0.33 |
| D15 | $3 \times 10^{-4}$ | $3.4 \times 10^{7}$ | D | SF | 97 | 640 | Weakly-rotating | 0.65 |
| D16 | $3 \times 10^{-4}$ | $3.4 \times 10^{7}$ | D | SF | 97 | 640 | Weakly-rotating | 0.98 |
| D17 | $3 \times 10^{-4}$ | $3.4 \times 10^{7}$ | D | SF | 97 | 640 | Weakly-rotating | 1.30 |
| D18 | $3 \times 10^{-4}$ | $3.4 \times 10^{7}$ | D | SF | 97 | 640 | Weakly-rotating | 1.63 |
| D19 | $3 \times 10^{-4}$ | $2.4 \times 10^{6}$ | D | SF | 97 | 640 | Moderately-rotating | 0.0 |
| D20 | $3 \times 10^{-4}$ | $2.4 \times 10^{6}$ | D | SF | 97 | 640 | Moderately-rotating | 0.33 |
| D21 | $3 \times 10^{-4}$ | $2.4 \times 10^{6}$ | D | SF | 97 | 640 | Moderately-rotating | 0.65 |
| D22 | $3 \times 10^{-4}$ | $2.4 \times 10^{6}$ | D | SF | 97 | 640 | Moderately-rotating | 0.98 |
| D23 | $3 \times 10^{-4}$ | $2.4 \times 10^{6}$ | D | SF | 97 | 640 | Moderately-rotating | 1.30 |
| D24 | $3 \times 10^{-4}$ | $2.4 \times 10^{6}$ | D | SF | 97 | 640 | Moderately-rotating | 1.63 |
| D25 | $1.5 \times 10^{-4}$ | $1.1 \times 10^{8}$ | D | SF | 97 | 864 | Weakly-rotating | 0.0 |
| D26 | $1.5 \times 10^{-4}$ | $1.1 \times 10^{8}$ | D | SF | 97 | 864 | Weakly-rotating | 0.33 |
| D27 | $1.5 \times 10^{-4}$ | $1.1 \times 10^{8}$ | D | SF | 97 | 864 | Weakly-rotating | 0.65 |
| D28 | $1.5 \times 10^{-4}$ | $1.1 \times 10^{8}$ | D | SF | 97 | 864 | Weakly-rotating | 0.98 |
| D29 | $1.5 \times 10^{-4}$ | $1.1 \times 10^{8}$ | D | SF | 97 | 864 | Weakly-rotating | 1.30 |
| D30 | $1.5 \times 10^{-4}$ | $1.1 \times 10^{8}$ | D | SF | 97 | 864 | Weakly-rotating | 1.63 |
| D31 | $1.5 \times 10^{-4}$ | $7.9 \times 10^{6}$ | D | SF | 97 | 864 | Moderately-rotating | 0.0 |
| D32 | $1.5 \times 10^{-4}$ | $7.9 \times 10^{6}$ | D | SF | 97 | 864 | Moderately-rotating | 0.33 |
| D33 | $1.5 \times 10^{-4}$ | $7.9 \times 10^{6}$ | D | SF | 97 | 864 | Moderately-rotating | 0.65 |
| D34 | $1.5 \times 10^{-4}$ | $7.9 \times 10^{6}$ | D | SF | 97 | 864 | Moderately-rotating | 0.98 |
| D35 | $1.5 \times 10^{-4}$ | $7.9 \times 10^{6}$ | D | SF | 97 | 864 | Moderately-rotating | 1.30 |
| D36 | $1.5 \times 10^{-4}$ | $7.9 \times 10^{6}$ | D | SF | 97 | 864 | Moderately-rotating | 1.63 |
| D37 | $6 \times 10^{-5}$ | $5.4 \times 10^{8}$ | D | SF | 97 | 1280 | Weakly-rotating | 0.0 |
| D38 | $6 \times 10^{-5}$ | $5.4 \times 10^{8}$ | D | SF | 97 | 1280 | Weakly-rotating | 0.33 |
| D39 | $6 \times 10^{-5}$ | $5.4 \times 10^{8}$ | D | SF | 97 | 1280 | Weakly-rotating | 0.65 |
| D40 | $6 \times 10^{-5}$ | $5.4 \times 10^{8}$ | D | SF | 97 | 1280 | Weakly-rotating | 0.98 |
| D41 | $6 \times 10^{-5}$ | $5.4 \times 10^{8}$ | D | SF | 97 | 1280 | Weakly-rotating | 1.30 |
| D42 | $6 \times 10^{-5}$ | $5.4 \times 10^{8}$ | D | SF | 97 | 1280 | Weakly-rotating | 1.63 |
| D43 | $6 \times 10^{-5}$ | $3.4 \times 10^{7}$ | D | SF | 97 | 1280 | Moderately-rotating | 0.0 |
| D44 | $6 \times 10^{-5}$ | $3.4 \times 10^{7}$ | D | SF | 97 | 1280 | Moderately-rotating | 0.33 |
| D45 | $6 \times 10^{-5}$ | $3.4 \times 10^{7}$ | D | SF | 97 | 1280 | Moderately-rotating | 0.65 |
| D46 | $6 \times 10^{-5}$ | $3.4 \times 10^{7}$ | D | SF | 97 | 1280 | Moderately-rotating | 0.98 |
| D47 | $6 \times 10^{-5}$ | $3.4 \times 10^{7}$ | D | SF | 97 | 1280 | Moderately-rotating | 1.30 |
| D48 | $6 \times 10^{-5}$ | $3.4 \times 10^{7}$ | D | SF | 97 | 1280 | Moderately-rotating | 1.63 |
| D49 | $3 \times 10^{-5}$ | $1.8 \times 10^{9}$ | D | SF | 129 | 1536 | Weakly-rotating | 0.0 |
| D50 | $3 \times 10^{-5}$ | $1.8 \times 10^{9}$ | D | SF | 129 | 1536 | Weakly-rotating | 0.98 |

| D51 | $3 \times 10^{-5}$ | $1.8 \times 10^{9}$ | D | SF | 129 | 1536 | Weakly-rotating | 1.63 |
| D52 | $3 \times 10^{-5}$ | $1.2 \times 10^{8}$ | D | SF | 129 | 1536 | Moderately-rotating | 0.0 |
| D53 | $3 \times 10^{-5}$ | $1.2 \times 10^{8}$ | D | SF | 129 | 1536 | Moderately-rotating | 1.63 |

           

\end{document}